\documentclass[11pt, fleqn]{article}

\usepackage{amsmath}
\usepackage{amssymb}
\usepackage{amsfonts}
\usepackage{amsthm}
\usepackage{dsfont}
\usepackage{mathrsfs}
\usepackage{ulem}

\usepackage{enumerate}

\usepackage[arrowdel]{physics}
\usepackage{tensor}

\usepackage{simplewick}
\usepackage{feynmf}

\usepackage{tabu}
\usepackage{physics}
\usepackage{mathdots}
\usepackage{diagbox}

\newcommand{\operator}[1]{\hat{#1}}

\newcommand{\R}{\mathbb{R}}

\newcommand{\Z}{\mathbb{Z}}

\newcommand{\RS}{\R^3\times\S^1}
\newcommand{\RT}{\R^2\times\T^2_*}
\newcommand{\RTS}{\R\times\T^2_*\times S^1}
\newcommand{\T}{\mathbb{T}}

\newcommand{\SU}[1]{SU(#1)}
\newcommand{\U}[1]{U(#1)}

\newcommand{\pauli}[1]{
    \ifnum#1=1
        \operator{\sigma}_{x}
    \else
        \ifnum#1=2
           \operator{\sigma}_{y}
        \else
            \ifnum#1=3
                \operator{\sigma}_{z}
            \else
                \errmessage{Incorrect number given to pauli}
            \fi
        \fi
    \fi
}

\usepackage{geometry}
\usepackage{comment}

\usepackage{soul}
\usepackage{subcaption}

\usepackage{jheppub}
\usepackage{amsmath,amssymb,amsthm}
\usepackage{bm}
\usepackage{slashed}
\usepackage{graphicx}
\usepackage[T1]{fontenc}
\usepackage{fix-cm}

\usepackage{tabu}
\usepackage{physics}

\makeatletter
\newcommand*\bigcdot{\mathpalette\bigcdot@{.5}}
\newcommand*\bigcdot@[2]{\mathbin{\vcenter{\hbox{\scalebox{#2}{$\m@th#1\bullet$}}}}}
\makeatother

\newlength{\dummysp}
\usepackage[font=scriptsize]{caption}
\usepackage{mwe}

\def\R{{\mathbb R}}
\def\S{{\mathbb S}}
\def\Z{{\mathbb Z}}
\def\T{{\mathbb T}}

\title{Numerical fractional instantons in $\SU{2}$: center vortices, monopoles, and a sharp transition between them}
\author{F. David Wandler}
\affiliation{Department of Physics, University of Toronto, Toronto, ON M5S 1A7, Canada}
\emailAdd{f.wandler@mail.utoronto.ca}

 \abstract{ We use a numerical cooling algorithm to study fractional instantons in $\SU{2}$ pure Yang-Mills on $\RT$, $\RS$, and $\RTS$. We confirm that the fractional instantons are center vortices on $\RT$ and monopoles on $\RS$, and we calculate several properties relevant to using these solutions for semiclassical calculations. On $\RTS$, we interpolate between the large $\T^2_*$ limit and the large $S^1$ limit to study how the solutions interpolate between center vortices and monopoles. We find that they are separated by a sharp transition, with 't Hooft's constant field strength solutions living at the transition point. These results contrast but do not contradict recent results suggesting continuity between vortices and monopoles. }

\begin{document}

\maketitle

\section{Introduction}

In the past two decades, considerable progress has been made in understanding the confinement of 4D non-abelian gauge theories through center-stabilized compactifications of Yang-Mills (YM) theory\footnote{The literature is now far too large to include every possible reference. Instead, we focus on the initial papers related to each model. We also recommend \cite{Poppitz:2021cxe} as a review of center-stabilization and confinement on $\RS$.}. These theories involve starting with 4D Yang-Mills theory with one or more compact (but finite) directions and the addition of a stabilizing feature to ensure that the center symmetry does not break. The preservation of center symmetry ensures the theory remains in the confining regime. Interestingly, these theories allow for semiclassical confinement mechanisms that rely on \textit{fractional instantons}, instantons with fractional topological charge.

For example, \textit{supersymmetric YM} (SYM), \textit{deformed YM} (dYM), and \textit{adjoint QCD} (QCD(adj)) can all be shown to confine on $\RS$ (with periodic fermions) through mechanisms involving monopole instantons \cite{Unsal2008adj,Unsal2009bions,Shifman:2008ja,Unsal:2008ch}. These theories all abelianize in the IR, and crucially, the monopole-instanton configurations have magnetic charges in the abelian EFT. The condensation of these charges (or pairs of charges called \textit{bions}) causes confinement.

Another example is pure YM on $\RT$. Here the ${}_*$ denotes the inclusion of an 't Hooft flux (i.e. twisted boundary conditions) on the $\T^2$. This theory was shown to abelianize to a $\Z_2$ gauge theory and confine due to center vortices \cite{Tanizaki:2022ngt}. For a more thorough discussion of ensembles of center vortices in 3D and 4D (including monopoles living on the center vortex worldvolumes), see \cite{Oxman:2018prd,Oxman2018epjc,Junior2020,Junior2021,Junior2022}.

While these theories are all shown to be in the confining phase, the confinement mechanisms are very different. This raises the natural question: \textit{to what extent are the different confinement mechanisms related and are the theories continuously connected?} 

In search of an answer, the author with Erich Poppitz wrote \cite{Poppitz2023}. In this paper, we presented a pair of classical vacua that should allow for a semiclassical understanding of continuity of dYM in the semi-infinite volume limit, $\RTS\rightarrow\RS$. We worked out that these vacua, which had been considered previously in \cite{Perez:2013aa} and \cite{Unsal:2020yeh}, had many desirable properties, such as abelianization of the gauge group and exchange under center symmetry. We then confirmed that they were stable to quantum fluctuations through calculating the one-loop potential on the holonomy. However, we did not explore the full semiclassical calculations leading to confinement on $\RTS$.

The main difficulty in extending \cite{Poppitz2023} to a full semiclassical analysis and opening up the entire landscape of $\RTS$ theories was a lack of understanding of the fractional instantons. The present article makes progress in understanding these fractional instanton solutions through numerics. In particular, we find lattice approximations of the fractional instanton solutions by minimizing the twisted $\SU{2}$ Wilson action \cite{Wilson:1974sk} using the cooling algorithm from \cite{Laursen1988,GarciaPerez:1989gt,GarciaPerez:1992fj,GONZALEZARROYO1998273}, which was adapted from the algorithms used in \cite{Berg:1981nw} and \cite{Teper:1985rb}.

We first confirm the existence of center vortices on $\RT$, strengthening the initial evidence from \cite{GONZALEZARROYO1998273} and supporting the confinement mechanism proposed in \cite{Tanizaki:2022ngt}. We then confirm the existence of monopole instantons on $\RS$, agreeing with analytical work in \cite{GPY1981,Unsal2008adj,Unsal2009bions,Shifman:2008ja,Unsal:2008ch}. 

Having confirmed the algorithm through its results in the two known regimes, we consider the fractional instantons on $\RTS$ as it interpolates between the large torus (i.e. $\RS$) and large circle (i.e. $\RT$) limits. This is done by running the cooling algorithm on lattices with different sizes. We use $N_i$ to denote the number of lattice sites in the $i$-direction. The $N_i$ are directly analogous to the lengths of the dimensions of the torus in the continuum limit. 

While holding $N_0$ fixed at a large value and varying the remaining $N_i$, we find that both monopoles and center vortices can exist, but that there is a sharp transition between lattices that permit center vortices and lattices that permit monopoles. Namely, given the dimensionless parameter 
\begin{equation}
    \Delta = \frac{N_1N_2-N_0N_3}{\sqrt{N_0N_1N_2N_3}}\, ,
\end{equation}
the type of fractional instanton present is determined by the sign of $\Delta$\footnote{As described later in the paper, we perform our simulations with fluxes in the $12$- and $03$-planes, so that the $12$-plane corresponds to the $\T_*$ in $\RTS$. The numerator of our $\Delta$ parameter reflects this structure.}. Center vortices are found whenever $\Delta$ is negative and monopoles whenever $\Delta$ is positive. This suggests a phase diagram-like structure for the parameter space, which is shown in Figure \ref{fig:phase_diagram}. 

These solutions can be directly applied to the confinement problem with varying degrees of success. The center vortices are known to cause confinement on $\RT$ through direct calculation of the area law of Wilson loops as presented in \cite{Tanizaki:2022ngt}. Here we find that, on $\RTS$ with small $\T^2_*$, they also cause confinement through instanton effects that preserve the center symmetry. This leads to a nice continuous understanding of the confinement mechanism through an entire landscape of theories along the $\RTS\rightarrow\RT$ limit. 

In contrast, it is known that the monopole instanton configurations are only useful for explaining confinement on $\RS$ in the presence of center stabilization (i.e. either adjoint periodic fermions or a non-local deformation potential). One limitation of the work presented here is that the basic cooling algorithm we employ is only suitable for studying pure YM, hence our findings do not directly provide a confinement mechanism. However, they are useful for understanding the properties of monopoles, in general. While there would be quantitative differences (such as differing monopole diameters) between our solutions and the solutions in dYM, we expect the same qualitative features between solutions, such as the shape of the surrounding magnetic field and the total magnetic flux. 

Another interesting limitation of the monopole solutions is that the $\Delta>0$ region of parameter space, which permits monopoles, becomes vanishingly small in the zero temperature limit (i.e. $N_0\rightarrow\infty$). This suggests that monopoles do not appear in pure YM on $\RTS$ at zero temperature, though they do appear at finite temperatures with large $\T^2_*$.

We also look at the "critical" case, $\Delta=0$. Here the lattice is self-dual and permits the constant field strength solutions of 't Hooft \cite{tHooft:1981nnx}. We find that our solutions on the lattice agree with these analytical solutions.

While preparing this article, the author became aware of \cite{hayashi2024unifying} and \cite{guvendik2024metamorphosis}. In contrast to the sharp transition apparent in our numerical results, these works provide evidence of a continuity between center vortices and monopole instantons. These results are not contradictory to ours since they consider different geometrical setup and/or theories differing from pure YM. The cooling algorithm presented here would be useful in investigating the claims in \cite{hayashi2024unifying}. A similar approach, but with an upgraded algorithm capable of handling the deformation potential, would be useful for interrogating the claims in \cite{guvendik2024metamorphosis} and for investigating confinement mechanisms on $\RTS$ with large $\T^2_*$. We discuss these results and the possibility of numerical investigations in greater detail in Sections \ref{sec:interpolation} and \ref{sec:conclusions}.

The data and code used in this study will be available on request starting September 2024.

\subsection{Layout}

This paper is organized as follows: Section \ref{sec:basics} outlines basics of the classical twisted Wilson action, the cooling algorithm we use to minimize it, and the notation that we use throughout the rest of the article. Section \ref{sec:general_features} discusses features that are true of all lattice sizes, such as the vacuum states without fractional instantons, the saturation of the BPS bound for all our final configurations, and the moduli space of the fractional instantons. Section \ref{sec:center_vortices} discusses the properties of fractional instantons in the $\R\times \T^2_*\times \R$ limit and confirms that they are center vortices. Similarly, Section \ref{sec:monopoles} discusses the properties of fractional instantons in the $\RS$ limit and confirms that they are monopoles. Section \ref{sec:interpolation} discusses the fractional instantons on $\RTS$ that interpolate between the $\R\times \T^2_*\times \R$ and $\RS$ limits. Its main result is the "phase" diagram in Figure \ref{fig:phase_diagram}, which illustrates the sharp transition we find between the center vortex solutions and the monopole solutions. The rest of Section \ref{sec:interpolation} is devoted to demonstrating the important features of solutions in each regime and at the transition point. Finally, Section \ref{sec:conclusions} wraps up the discussion and presents some interesting directions for future study.

\section{Lattice basics, notation, and conventions}
\label{sec:basics}
\subsection{Notation}
In this work, we focus on 4-dimensional square lattices with periodic boundary conditions. We label the directions 0, 1, 2, and 3, and use $N_i$ to denote the number of sites in each direction. Throughout this article, we investigate a variety of lattice sizes, which we present in the format: $N_0\times N_1\times N_2\times N_3$. 

While the Euclidean lattice action treats all directions the same, we find it useful to consider the $0$ direction as the Wick rotation of the time direction. It is purposely always taken to be the largest dimension, since we are interested in approximating the zero temperature limit: $\R\times T^2_*\times S^1$.

Our lattice sites are labelled by $x \in \Z_{N_0} \times \Z_{N_1} \times \Z_{N_2} \times \Z_{N_3}$, hence all arithmetic involving lattice sites is done modulo the number of sites in each direction. For the purposes of this arithmetic, we introduce $e_\mu \in \Z_{N_0} \times \Z_{N_1} \times \Z_{N_2} \times \Z_{N_3}$ as the unit vector in the $\mu^\text{th}$ direction. Individual components of $x$ are denoted by $x^\mu$ for $\mu=0,1,2,3$.

The link variable on the link extending in the positive $\mu$ direction from site $x$ is denoted by $U_\mu(x)\in \SU{2}$. Plaquette variables are defined as
\begin{equation}
    \square_{\mu\nu}(x) = U_\mu(x)U_\nu(x+e_\mu)U^\dag_\mu(x+e_\nu)U^\dag_\nu(x)\, .
\end{equation}

The action we consider is a twisted version of the Wilson action \cite{Wilson:1974sk}:
\begin{equation}
    S = \frac{1}{2g^2} \sum_{x} \sum_{0\leq \mu<\nu \leq 3} \Tr\left(1 - (-1)^{t_{\mu\nu}(x)} \square_{\mu\nu}(x) \right) \, .
\end{equation}
For the purposes of minimizing this classical action, the value of $g$ is irrelevant, so we set it to 1 for convenience.

Here $t_{\mu\nu}(x)$ is a binary value that indicates the presence of a twist in the $\mu\nu$ plaquette at $x$. If $t_{\mu\nu}$ is 0 everywhere, then there is no twist in the system and the action is the usual Wilson action. Since we are interested in understanding the system with an 't Hooft flux through the $12$-plane, we introduce a single twist in every $12$-plane in the system. In other words, $t_{\mu\nu}(x) = 1$ if $\mu=1$, $\nu=2$ and $x^1=x^2=0$. This is equivalent to introducing 't Hooft's twisted boundary conditions with $n_{12}=1$. 

When inspecting configurations that approximate continuum configurations with integer topological charge, such as the vacuum states for $\RTS$, we do not introduce any additional twists (i.e. all other $t_{\mu\nu}$ are set to 0). However, the fractional instanton configurations that we wish to investigate have fractional topological charge. Topological charge in the presence of twists take values in the set $\varepsilon^{\mu\nu\rho\lambda} \frac{n_{\mu\nu}n_{\rho\lambda}}{2} + \Z$. Thus, to ensure we approximate fractional topological charges, we must also include a twist in the $03$-plane. To accomplish this we keep the twists in plaquettes in the $12$-plane the same as above and set $t_{\mu\nu}(x)=1$ if $\mu=0$, $\nu=3$, and $x^0=x^3=0$. All other $t_{\mu\nu}(x)$ are set to 0. This is equivalent to boundary conditions with $n_{12}=n_{03}=1$ and all other components vanishing, leading to topological charges in $\frac{1}{2}+\Z$.

Given a topological charge, $Q_{top}$, in the continuum, the Bogomol'nyi-Prasad-Sommerfield (BPS) bound \cite{Bogomolny:1975de,Prasad:1975kr} is given by
\begin{equation}
\label{eqn:BPS_bound}
    S_{cont} \geq \frac{8\pi^2}{g^2} \left|Q_{top}\right| \, .
\end{equation}
We find that our lattice approximations of minimal continuum configurations with given topological charge\footnote{Here I am being very careful about the fact that the lattice configurations do not have nice cleanly defined topological charges like the continuum configurations they are approximating. In the rest of this article, I will be more relaxed with this language, and leave this footnote to avoid confusion.} also approximately saturate the BPS bound. This statement will be made more precise in Section \ref{sec:BPS_bound}, but it is a good test that our approach is making meaningful approximations of the continuum theory.

For analyzing our classical configurations we must use gauge invariant quantities\footnote{It is possible to gauge fix everything and compare configurations directly, but this is computationally complicated and does not allow for an easy comparison between differently sized lattices or with the continuum limit.}. On the lattice, all gauge invariant quantities are traces of Wilson loops. Certain Wilson loops will be very important in the following sections, so we introduce special notation for them. The first examples are the plaquette variables $\square_{\mu\nu}(x)$ defined above. These are useful because $\square_{\mu\nu}(x)\sim e^{ia^2 F_{\mu\nu}(x)}$, where $a$ is the lattice spacing and $F^{\mu\nu}$ is the standard YM field strength tensor in the continuum limit. Hence the plaquette variables give us access to local gauge invariant information. However, local information is not sufficient to tell us everything about the system; in particularly, it is useful to consider the global information contained in loops that wind around our torus. These holonomies are defined as
\begin{equation}
    W_\mu(x) = U_\mu(x)U_\mu(x+e_\mu) U_\mu(x+2e_\mu)\cdots U_\mu(x + (N_\mu-1)e_\mu) \, .
\end{equation}
Recall that addition of the lattice site vectors is always done modulo the lattice size in each direction.

\subsection{The cooling algorithm}
The cooling algorithm employed in this study is based on the algorithm used by Gonzalez-Arroyo, Garcia Perez, Soderberg, and Montero in \cite{GarciaPerez:1989gt,GarciaPerez:1992fj,GONZALEZARROYO1998273}. An earlier study \cite{Laursen1988} by Laursen and Schierholz also used a version of this algorithm to find monopole configurations in pure YM without twists\footnote{Due to the lack of twists, these monopoles would eventually vanish under continued cooling, but they survived long enough to confirm their existence in the $\SU{2}$ and $\SU{3}$ vacua.}. 

The algorithm works by considering a single link variable at a time and giving it the value that minimizes the action while keeping all other links fixed. This procedure is then repeated for every link in the lattice. Minimizing each link variable once is known as a \textit{sweep}. Sweeps are repeated until a suitable convergence criteria is met. 

\begin{figure}
    \centering
    \includegraphics[width=0.7\textwidth]{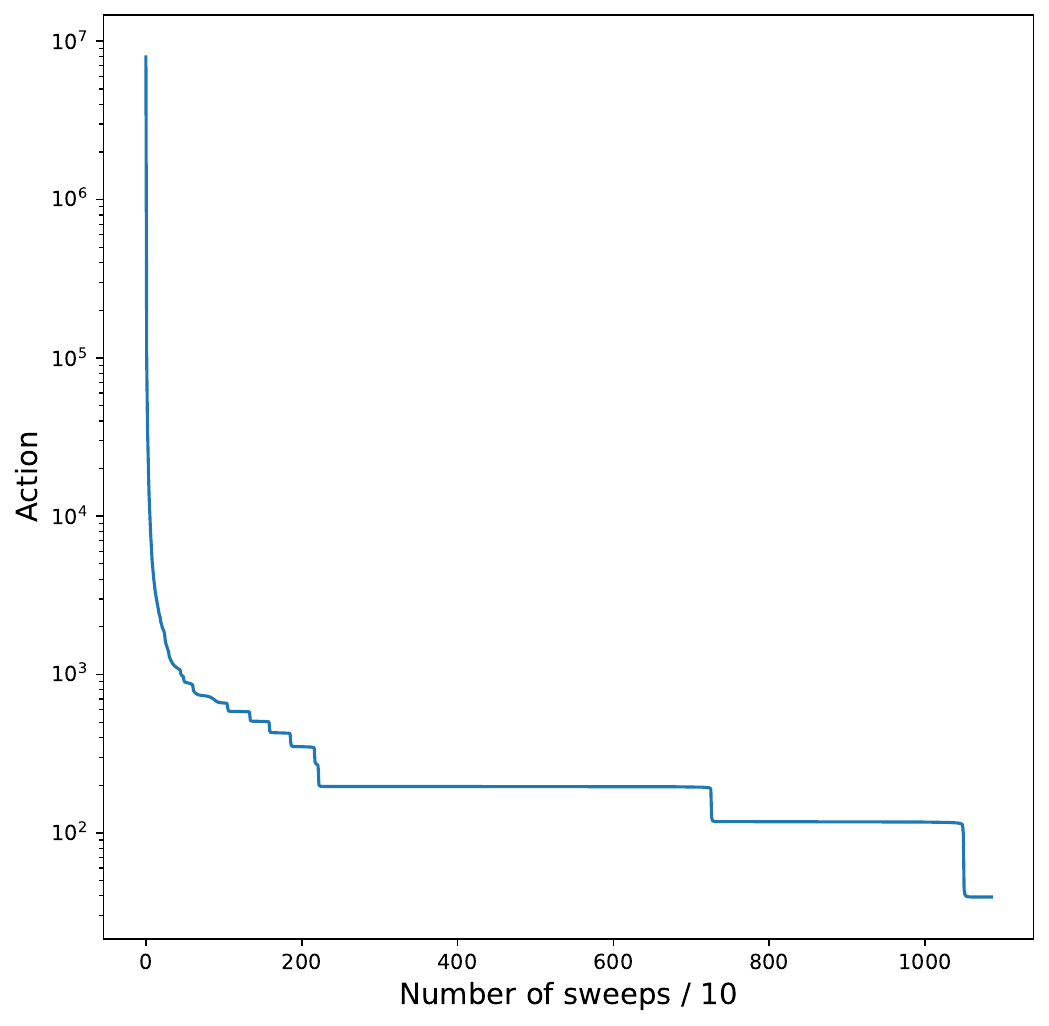}
    \caption{The action of the configuration as a function of the number of sweeps of the cooling algorithm. This particular example is for the lattice configuration $48\times48\times48\times12$.}
    \label{fig:action_vs_sweeps}
\end{figure}

In our case, the convergence criteria used is that the action changes by less than 1 ppm for 100 sweeps. We also require that the final value of the action is lower than 40. This was a practical measure to ensure the algorithm would not finish prematurely, since there is a tendency for the algorithm to sit in local minima for many, many sweeps before finding the true minima. This is demonstrated in the step-like pattern in Fig. \ref{fig:action_vs_sweeps} which plots the action as a function of the number of sweeps performed for a particular run. The height of the plateaus seems in line with the local minima being integer multiples of the instanton action above the BPS bound. This suggests that, as the algorithm minimizes, it "gets stuck" at states that are the desired ground state plus some configuration of instantons/antinstantons. In all cases, however, the action eventually settled to near the BPS bound.

The key feature of the algorithm that allows it to leave the local minima seems to be the fact that the link variables are replaced by their global minima (given the local neighbourhood). This can be done efficiently because of two features unique to $\SU{2}$. To see how this works consider a specific link variable, $U\in \SU{2}$. This link variable enters the action through the trace of six plaquettes. Hence the contribution of $U$ to the action that we want to minimize can be written as 
\begin{equation}
\label{eqn:local_action}
-\sum_{j=1}^{3} \Tr\left(U V_j^\dag\right) -\sum_{j=4}^{6} \Tr\left(U^{\dag} V_j\right)\,,
\end{equation}
where we have ignored an overall positive factor because it does not affect the where the minimum occurs. The $V_j$ are the combined products of the rest of the links making up the plaquettes. In the case of $\SU{2}$, the traces are always a real number, hence for all $A\in \SU{2}$,
\begin{equation}
\Tr\left(A^\dag\right) = \Tr\left(A\right)^* = \Tr\left(A\right)\,
\end{equation}
Thus we can rearrange \ref{eqn:local_action} to 
\begin{equation}
    -\sum_{j=1}^{6} \Tr\left(U V_j^\dag\right) = -\Tr\left(U \sum_{j=1}^{6}V_j^\dag\right)\,.
\end{equation}
Now we use the property that any sum of $\SU{2}$ matrices can be written as an $\SU{2}$ matrix times a non-negative real number, to write 
\begin{equation}
    \sum_{j=1}^{6}V_j^\dag = r W^\dag\,,
\end{equation}
where $W\in \SU{2}$ and $0<r\in \R$. It is now clear that we need to minimize
\begin{equation}
    -r\Tr\left(U W^\dag\right)\,.
\end{equation}
The minimum is $-r$ and is reached precisely when $U=W$. $W$ can be calculated from the surrounding links using standard linear algebra, so the algorithm replaces $U$ with $W$ then repeats this calculation for every link in the lattice to complete a sweep. 

\subsection{Data generated and analysis techniques}
With the intention of doing statistics on the space of configurations, 300 configurations were generated for three lattice shapes in both the $n_{03}=0$ and $n_{03}=1$ cases. The chosen lattice configurations were $24\times6\times6\times24$, $24\times6\times6\times6$, and $24\times24\times24\times6$, which correspond to the interesting cases $\RT$, $\R^1\times\T^3$, and $\RS$, respectively. These are used to draw a statistically meaningful bound on the size of parameter space that could cool to points outside of the observed moduli spaces. In addition, a wide variety of configurations with intermediate shapes and increased sizes were also considered in order to deduce the effects of the finite lattice spacing and to get a more fine grained understanding of the fractional instanton solutions. Due to computational constraints\footnote{These computational constraints were due to the fact that these calculations were performed on the author's personal computer, and could be easily overcome in follow-up studies with the use of dedicated scientific computing resources.}, these solutions were not generated in large numbers for each configuration. Instead, we treat them as examples demonstrating the properties of the individual instantons.

\begin{figure}
    \centering
    \includegraphics[width=\textwidth]{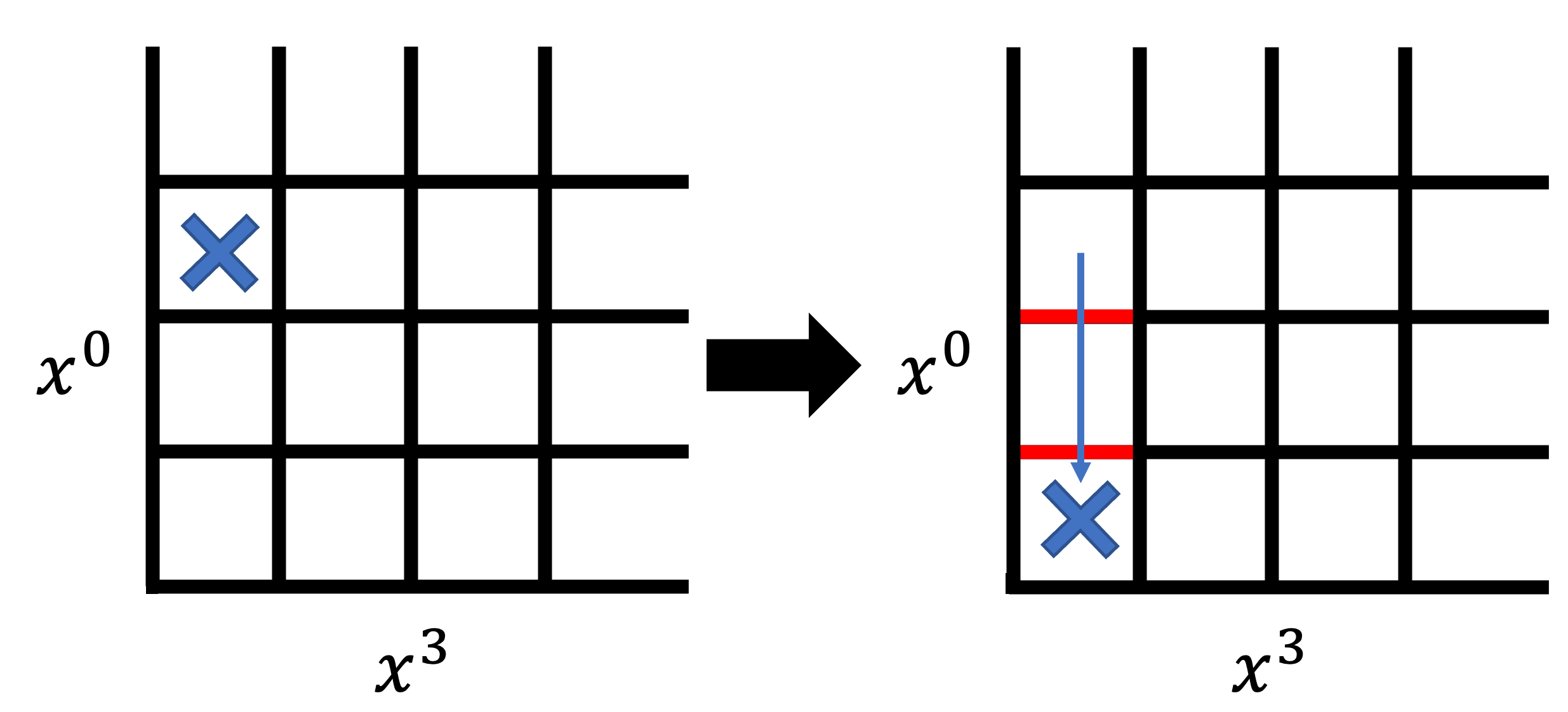}
    \caption{Cartoon example of the procedure for translating a twist in the $03$-plane along the $x^0$ direction. The twist is marked by the blue "X". From its initial location, we want to translate it down two plaquettes. The procedure is to multiply the two red links (the links crossed by the twist during the translation) by a $-1$.}
    \label{fig:translation}
\end{figure}

To directly compare the solutions, it is useful to translate the fractional instantons generated by different runs to align them to the same position on the torus. Translations by $n$ spaces in the $i^\text{th}$ direction are done in two steps. The first is to directly move all the values of the link variables by $n$ lattice sites in the $i$-direction. This translates the configuration, but also translates the location of the nonvanishing twists. To accurately compare the configurations, the second step is to translate the twists back to their original position. For a non-vanishing twist in the $ij$-plane, it is translated $n$ plaquettes back along the $i$-direction by multiplying all the link variables in the $j$-direction between the starting and ending plaquettes by $-1$. See Figure \ref{fig:translation} for a cartoon depicting this. 

Since there are twists in the $12$-plane, any translation in the $1$-direction will result in some $2$-direction links being multiplied by $-1$, and vice versa. The same will occur within the $03$-plane, when the $03$-twists are included. One consequence of this translation procedure is that if there is a twist in the $ij$-plane and the configuration is translated by $N_i$ in the $i$-direction, then the first step will not change the configuration, but the procedure for translating the twist will result in all the $j$-direction link variables with a single value of $x^i$ flipping sign. This is exactly a center symmetry transformation for loops winding around the $j$-direction. Hence, on the twisted torus, center symmetry transformations are equivalent to translations. This equivalence was previously reported in \cite{GarciaPerez:1992fj}.

This is a very useful fact in considering the moduli space of solutions. For example, it is known on $\RS$ that there are two distinct types of monopole-instantons, the BPS and KK monopoles. These instantons are however related by a center symmetry transformation of loops winding the $3$-direction, so in our setup they should be connected by translations by $N_0$ in the $0$-direction. Therefore, the distinct configurations in the infinite volume continuum become part of a connected moduli space on our periodic lattice.

\section{General features of the solutions}
\label{sec:general_features}
This section is devoted to features of the $\SU{2}$ solutions with $n_{12}=1$ that are not found to differ between different regions of parameter space, namely the properties of the vacuum when $n_{03}=0$, the saturation of the BPS bound for fractional instanton solutions with $n_{03}=1$, and the moduli space of the instanton solutions.

\subsection{True vacua}
\label{sec:true_vacua}
\begin{figure}
    \centering
    \includegraphics[width=1.0\textwidth]{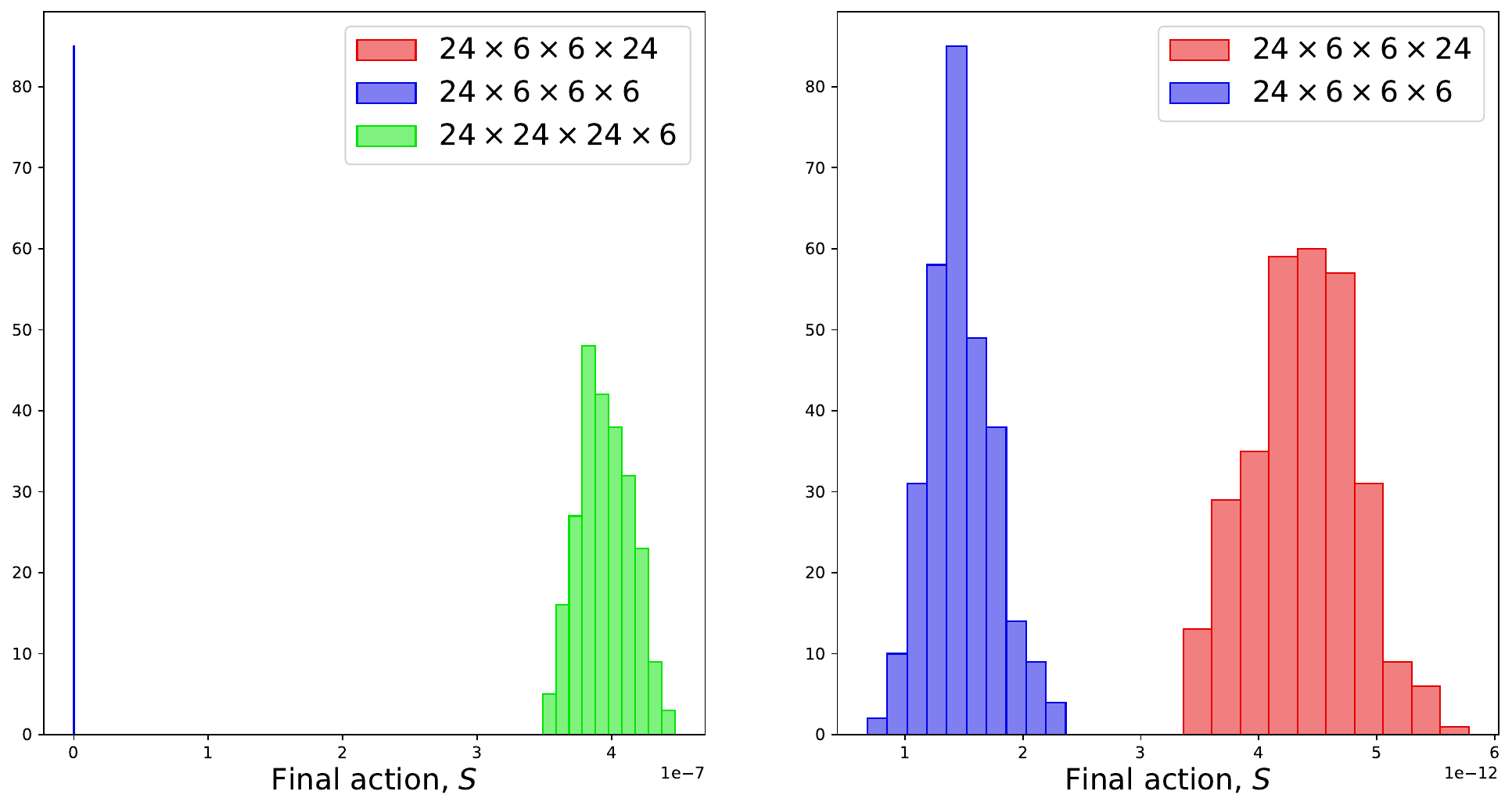}
    \caption{Histogram of the final actions for minimal configurations with $n_{03}=0$. The left plot includes all three datasets. The right plot zooms in on just the $24\times6\times6\times24$ and $24\times6\times6\times6$ datasets, as they overlap and are more difficult to see in the left plot.}
    \label{fig:untwisted_actions}
\end{figure}

Upon removing the 03-twist from any torus (i.e. setting $n_{03}=0$, but keeping $n_{12}=1$), the BPS bound from equation \ref{eqn:BPS_bound} is $S \geq 0$. The algorithm finds four discrete classical vacua that saturate this bound to within numerical error. To build more numerical support for this claim, 300 solutions were generated at various torus sizes. Figure \ref{fig:untwisted_actions} shows how torus shape affect the final action value. Notice in particular how the action stays below the precision target of $1.0\times10^{-6}$ (i.e. 1 ppm) for all configurations. This suggests that the algorithm is converging towards a true minimal action configuration that saturates the BPS bound. 

\begin{figure}
    \centering
    \includegraphics[width=0.9\textwidth]{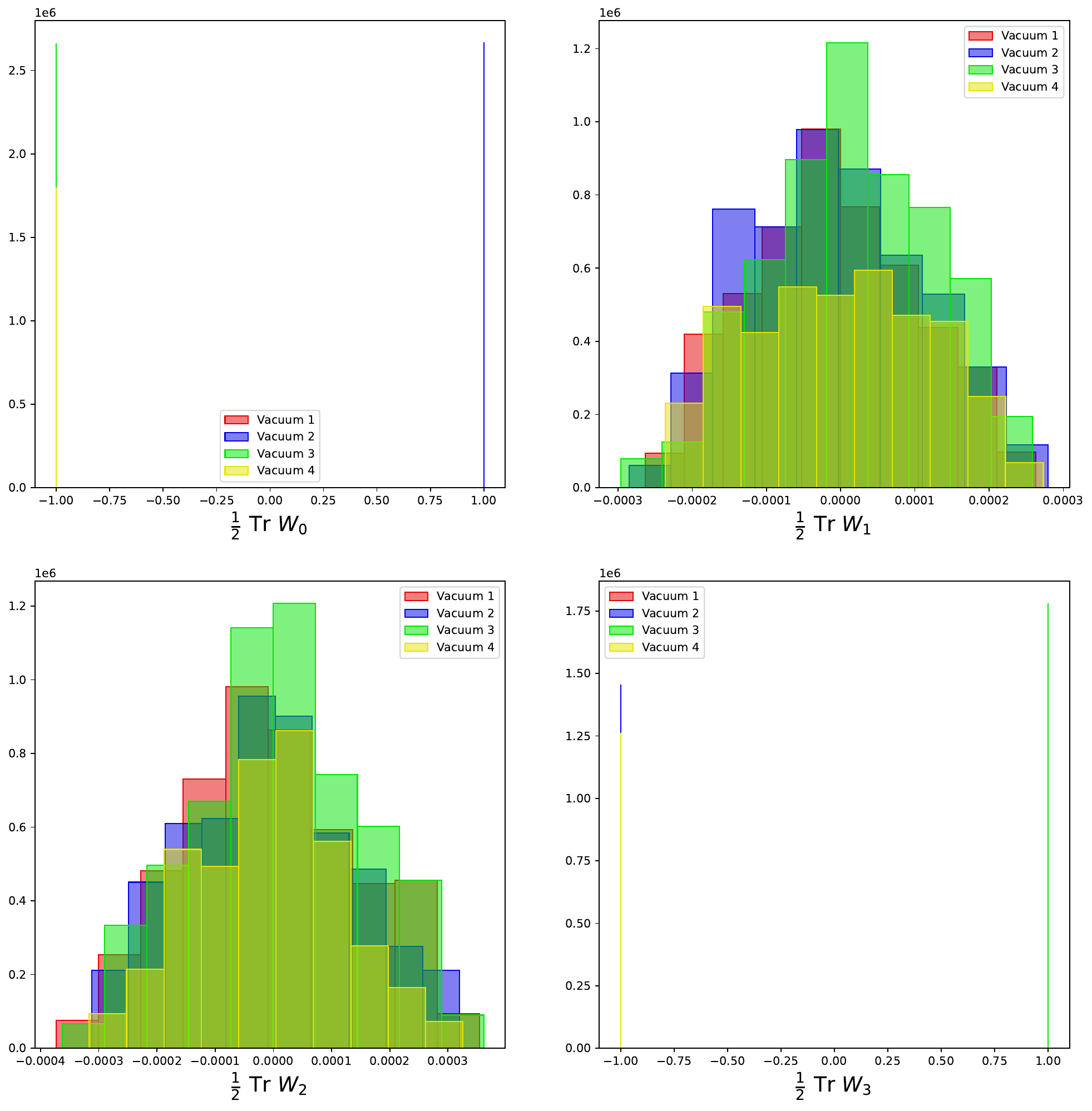}
    \caption{Histograms of the traces of the 4 holonomies from every lattice site across the $24\times24\times24\times6$ dataset with $n_{03}=0$.}
    \label{fig:untwisted_holos_1}
\end{figure}

\begin{figure}
    \centering
    \includegraphics[width=0.9\textwidth]{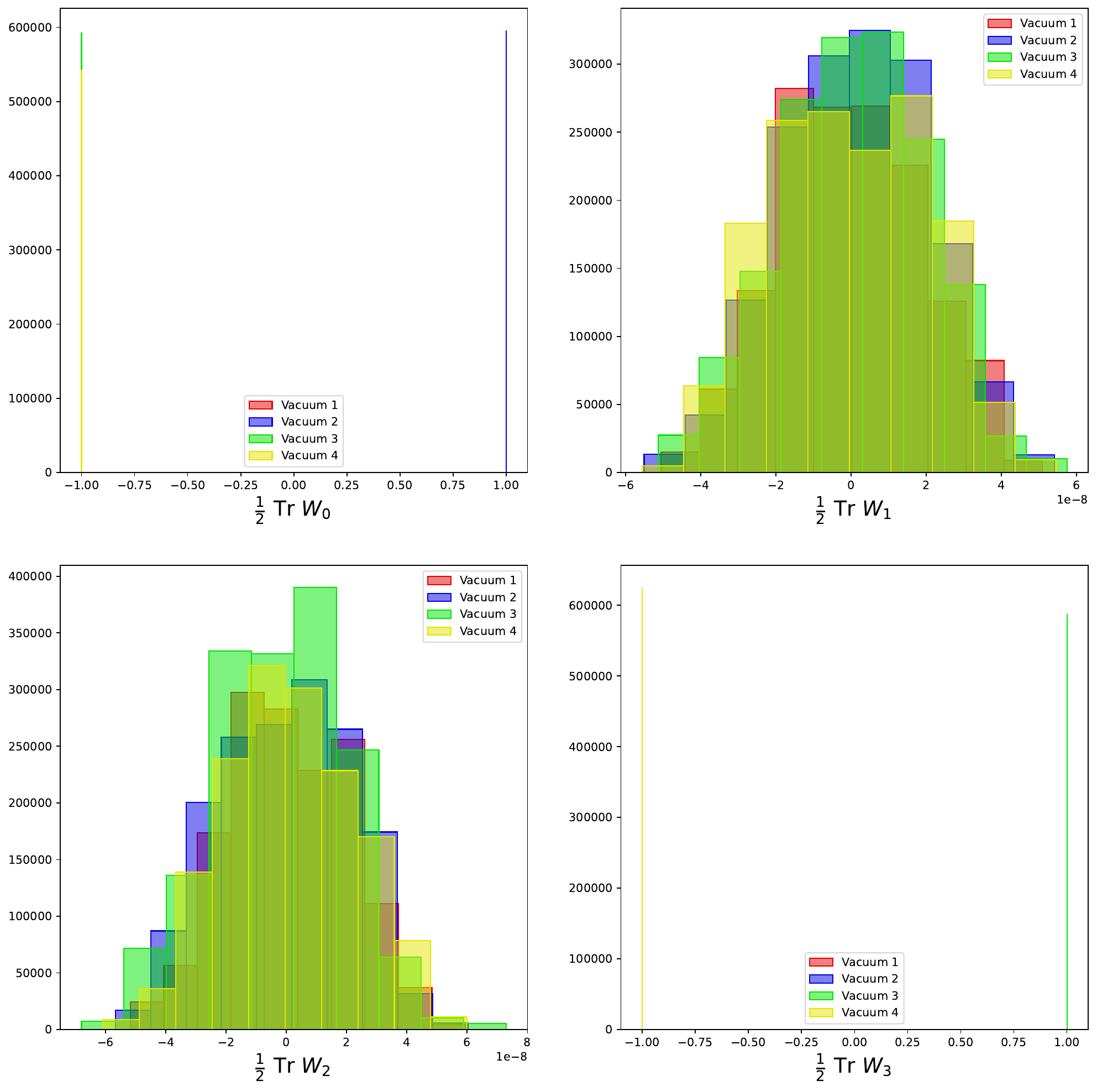}
    \caption{Histograms of the traces of the 4 holonomies from every lattice site across the $24\times6\times6\times24$ dataset with $n_{03}=0$.}
    \label{fig:untwisted_holos_2}
\end{figure}

\begin{figure}
    \centering
    \includegraphics[width=0.9\textwidth]{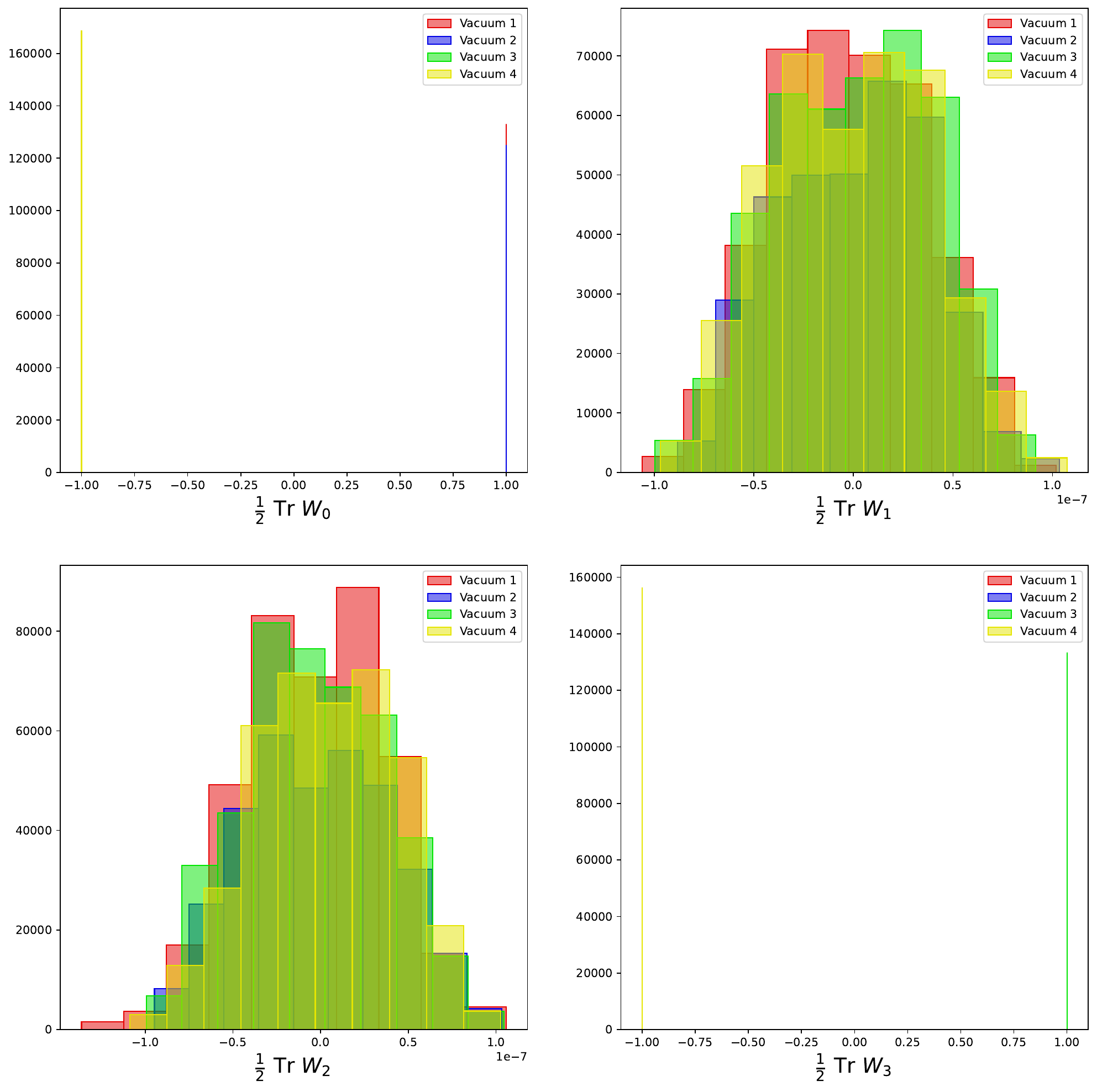}
    \caption{Histograms of the traces of the 4 holonomies from every lattice site across the $24\times6\times6\times6$ dataset with $n_{03}=0$.}
    \label{fig:untwisted_holos_3}
\end{figure}

The fact that there are four such vacua can be seen from the measured holonomies at $n_{12}=1$ and $n_{03}=0$. The holonomies in each direction are constant across the entire lattice and are given (up to numerical noise) by
\begin{equation}
\label{eqn:vacua}
    \begin{split}
    \frac{1}{2}\Tr W_1 = & \,0 \\
    \frac{1}{2}\Tr W_2 = & \,0 \\
    \frac{1}{2}\Tr W_3 = & \pm 1 \\
    \frac{1}{2}\Tr W_0 = & \pm 1
    \end{split}
\end{equation}
Here the value of $\Tr W_3$ is independent of the value of $\Tr W_0$, leading to the four distinct states. This counting of vacua is consistent with \cite{Gonzalez-Arroyo:1997ugn} where it is shown that $\SU{N}$ has $N^2$ vacua. Our findings are also consistent with the center symmetry in the 0- and 3-directions being spontaneously broken, while the center symmetry in the 1- and 2- directions is unbroken. To see that this holds generally, Figures \ref{fig:untwisted_holos_1}, \ref{fig:untwisted_holos_2}, and \ref{fig:untwisted_holos_3} give histograms of the values of $\Tr W_i$ for $i=1,2,3,4$ which include values from every lattice site for all 300 configurations at each size. The four vacua are determined by the signs of $W_0$ and $W_3$, as detailed in Table \ref{tab:vacua}. The narrow peaks around the exact values mentioned above suggest that there are no alternative vacua and that the vacua are consistent across various lattice shapes.  

\begin{table}
    \centering
    \begin{tabular}{|c|cc|}
    \hline
         Vacuum & $W_0$ & $W_3$ \\
    \hline
         1 & + & + \\
         2 & + & - \\
         3 & - & + \\
         4 & - & - \\
    \hline
    \end{tabular}
    \caption{Signs of $W_0$ and $W_3$ associated with the 4 distinct vacua in Figures \ref{fig:untwisted_holos_1}, \ref{fig:untwisted_holos_2}, and \ref{fig:untwisted_holos_3}.}
    \label{tab:vacua}
\end{table}

Another interesting gauge invariant in these solutions is $\Tr W_1W_2$, where the base points of the $W_1$ and $W_2$ coincide. Recall that, as $\SU{2}$ matrices, each $W_i$ can be written as $\cos(\theta_i) + i \sin(\theta_i) \hat{n}\cdot\vec{\sigma}$. In the case where $\Tr W_i=0$, we know that $\cos(\theta_i)=0$ and $\sin(\theta_i)=1$, hence $W_i=\hat{n}_i\cdot\vec{\sigma}$. Therefore, for these vacua solutions we have $\Tr W_1W_2\propto \hat{n}_1\cdot\hat{n}_2$. Calculating $\Tr W_1(x)W_2(x)$ for every lattice, $x$, in every configuration from the large datasets, we find the data in Table \ref{tab:untwisted_W1W2}. It is clear that our numerical solutions are consistent with $\Tr W_1W_2=0$.

\begin{table}
    \centering
    \begin{tabular}{|c|ccc|}
        \hline
         Dataset & Mean $\Tr W_1W_2$ & Minimal $\Tr W_1W_2$ & Maximal $\Tr W_1W_2$ \\
         \hline
         $24\times6\times6\times24$ & $0.0013\times10^{-12}$ & $-1.4163\times10^{-12}$ & $1.3476\times10^{-12}$ \\
         $24\times6\times6\times6$ & $-0.0055\times10^{-13}$ & $-6.9555\times10^{-13}$ & $7.1887\times10^{-13}$ \\
         $24\times24\times24\times6$ & $0.0022\times10^{-6}$ & $-1.9887\times10^{-6}$ & $2.3290\times10^{-6}$ \\
         \hline
    \end{tabular}
    \caption{Values of $\Tr W_1W_2$ calculated across the entirety of the large datasets.}
    \label{tab:untwisted_W1W2}
\end{table}

In the case where $L_i \ll \Lambda$ and $\frac{1}{2}\Tr W_i \neq \pm 1$, we have that the vev of $W_i$ will higgs the gauge symmetry down to a $\U{1}$ subgroup which aligns with $\hat{n}_i$ in color space. The strongest breaking occurs for $\frac{1}{2}\Tr W_i=0$. Hence, in cases where both $L_1,L_2 \ll \Lambda$, these vacua are maximally break $\SU{2}$ into two orthogonal $\U{1}$ subgroups. This suggests that the gauge group is actually broken to the $\Z_2$, since the $\Z_2$ center of $\SU{2}$ is the only possible intersection between orthogonal $U(1)$ subgroups. Hence, the low energy theory around these vacua is a $\Z_2$ gauge theory without propagating degrees of freedom. 

All of these results for the vacua are in agreement with the known vacuum solutions in the continuum. Consider a continuum Euclidean torus with constant boundary conditions ($\Omega_0=\Omega_3=1$, $\Omega_1=\pauli{2}$, and $\Omega_2=\pauli{3}$), then the solutions $A=0$ and $A=-iT_3dT_3^{-1}$ are classical solutions with vanishing action\footnote{Here $T_3$ is the large gauge transformation implementing a center symmetry transformation on loops winding in the 3-direction.}. One can directly check that the gauge invariants all match with the numerical solutions described above. The agreement between the numerical lattice solutions and the analytical continuum solutions suggests that the numerical methods employed in this paper are a reliable approach to extracting information about classical Yang-Mills solutions. 

\subsection{Saturation of the BPS bound}
\label{sec:BPS_bound}
As seen above, the vacuum states saturate the BPS bound for the case of $Q_{top}=0$, where the BPS bound is $S_{Q_{top}=0} \ge 0$. A similar analysis shows that the BPS bound is saturated for the fractional instanton configurations with $Q_{top}=1/2$, where the BPS bound is $S_{Q_{top}=1/2} \ge 4\pi^2 / g^2$. As mentioned above, we have set $g=1$ for finding classical solutions, so we expect the action of our fractional instantons to be $S = 4\pi^2 \approx 39.48$. 

\begin{figure}
    \centering
    \includegraphics[width=0.9\textwidth]{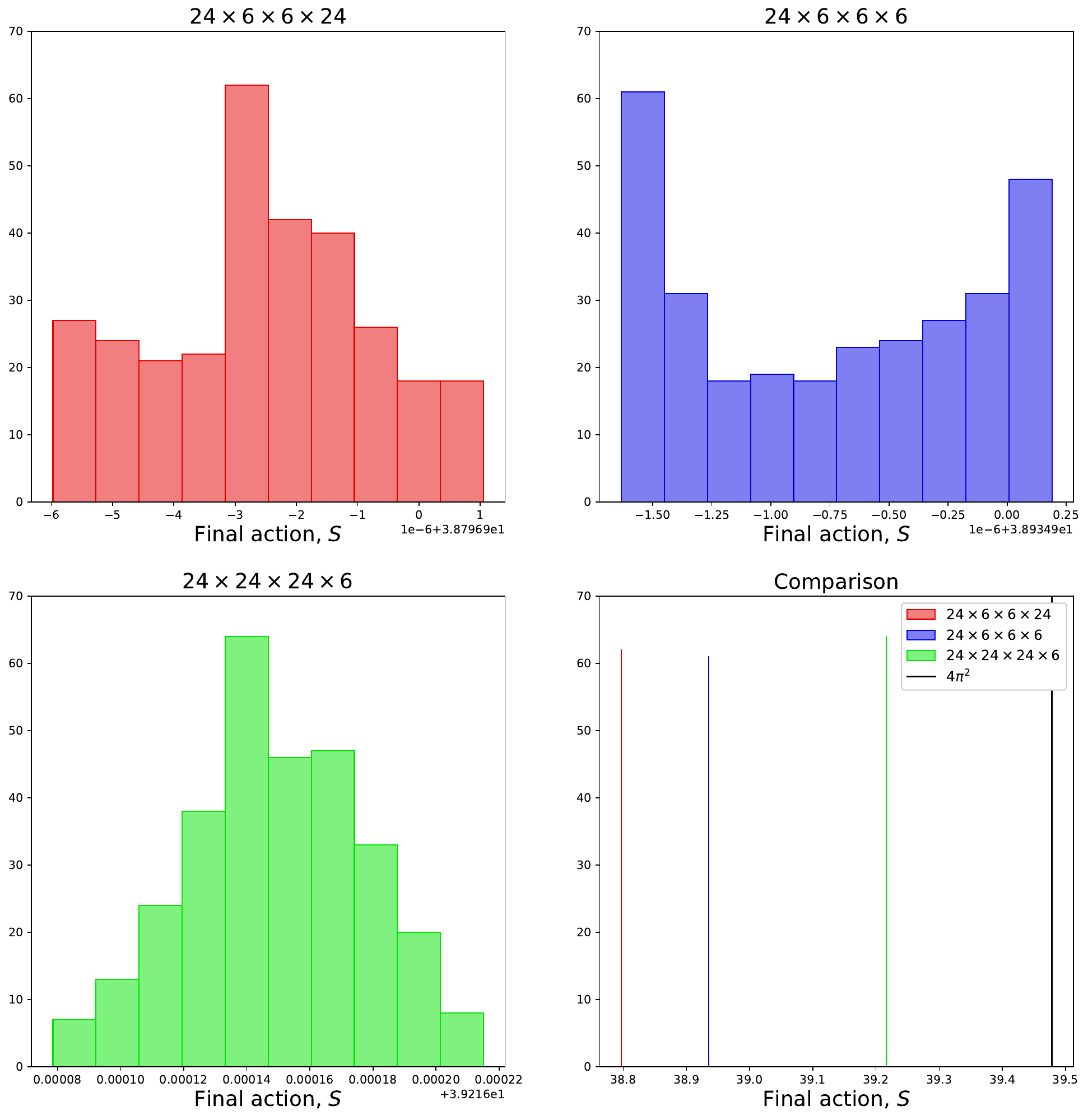}
    \caption{Histograms showing the final actions of all configurations with in the three datasets with $n_{03}=1$.}
    \label{fig:twisted_actions}
\end{figure}

Figure \ref{fig:twisted_actions} compares the actions of 300 configurations with $n_{03}=1$ at different lattice shapes. It is clear that the values are close to the expected value, but do not hit it exactly. Following \cite{GONZALEZARROYO1998273}, we can demonstrate that this discrepancy is a lattice effect by finding the effect of varying the lattice sizes on the action. We focus on the three cases $N_L\times N_L\times N_L\times N_S \sim \RS$, $N_L\times N_S\times N_S\times N_S\sim \RTS$, and $N_L\times N_S\times N_S\times N_L\sim \RT$, where $N_L$ is the "large" value and $N_S$ is the "small" value. The results are collected in Table \ref{tab:actions}. Given the 1 ppm convergence criteria, the values are expected to have numerical uncertainty on the order of $4\times10^{-5}$, so we present the values with 5 decimal places. When $N_L=24$, solutions are relatively quick to generate, so we consider the average over 300 (for $N_S=6$) or 10 (for $N_S=9,12$) solutions. Comparing the minimum and maximum final actions to the mean, we find a that the largest deviation is less than $8\times10^{-5}$. For larger values of $N_L$, we consider only a single solution. 

\begin{table}
\centering
\begin{tabular}{|c|c c c|}
    \hline
     & \multicolumn{3}{c|}{$N_L\times N_S \times N_S \times N_L$} \\
     \hline
     \diagbox{$N_S$}{$N_L$} & 24 & 36 & 48 \\
     \hline
     6 & $38.79690 \substack{+4\times10^{-6}\\-3\times10^{-6}}$ & 38.79688 & 38.79688 \\
     9 & $39.18044 \substack{+5\times10^{-8}\\-7\times10^{-8}}$ & 39.17992 & 39.17991 \\
     12 & $39.31421 \pm2\times10^{-6}$ & 39.31145 & 39.31136 \\
     \hline
     \multicolumn{4}{c}{}\\
     \hline
     &  \multicolumn{3}{c|}{$N_L\times N_S \times N_S \times N_S$}\\
     \hline
     \diagbox{$N_S$}{$N_L$} & 24 & 36 & 48 \\
     \hline
     6 & $38.93490 \pm9\times10^{-7}$ & 38.93486 & 38.93486\\
     9 & $39.24188 \substack{+3\times10^{-7}\\-5\times10^{-7}}$ & 39.24081 & 39.24079 \\
     12 & $39.35070 \substack{+4\times10^{-6}\\-6\times10^{-6}}$ & 39.34574 & 39.34554 \\
     \hline
     \multicolumn{4}{c}{}\\
     \hline
     & \multicolumn{3}{c|}{$N_L\times N_L \times N_L \times N_S$}\\
     \hline
     \diagbox{$N_S$}{$N_L$} & 24 & 36 & 48 \\
     \hline
     6 & $39.21615 \pm0.00007$ & 39.23172 & 39.23967\\
     9 & $39.35266 \substack{+0.00004\\-0.00006}$ & 39.36367 & 39.36934\\
     12 & $39.40310 \substack{+0.00005\\-0.00004}$ & 39.41032 & 39.41520\\
     \hline
\end{tabular}
\caption{Values of the final actions for different sizes and shapes of lattices. Values with errors are averaged across a dataset of multiple runs at the same lattice size. Errors are determined by the maximum and minimum values across the dataset. }
\label{tab:actions}
\end{table}

By changing the values of $N_L$ and $N_S$ separately, it is clear that the dependence on $N_S$ is far stronger for all cases. The observed weak dependence on $N_L$ is desirable, since we want our large dimension to be a good approximation of infinite length limit. We want to quantify the lattice effects on the minimal action. This would normally be done through the lattice spacing, $a$, but in the classical case there are no physical scales that can be used to meaningfully set the value of $a$. Any lattice spacing can be set for any size of lattice and the solutions will not change, so just taking the $a\rightarrow\infty$ limit is not useful. Instead, the limit we wish to understand is the limit of infinite lattice sites in all directions where the length of the larger dimensions tends towards infinity while the length of the smaller directions remains finite. In other words, we want to consider $N_L,N_S\rightarrow \infty$ such that $aN_L\rightarrow\infty$ and $aN_S\rightarrow L$ for some finite $L$. Hence, the lattice spacing should scale like $L/N_S$. Since we are free to normalize distance scales to  $L=1$, $1/N_S$ as a good dimensionless substitute for the lattice spacing.

\begin{figure}
    \centering
    \includegraphics[width=0.8\textwidth]{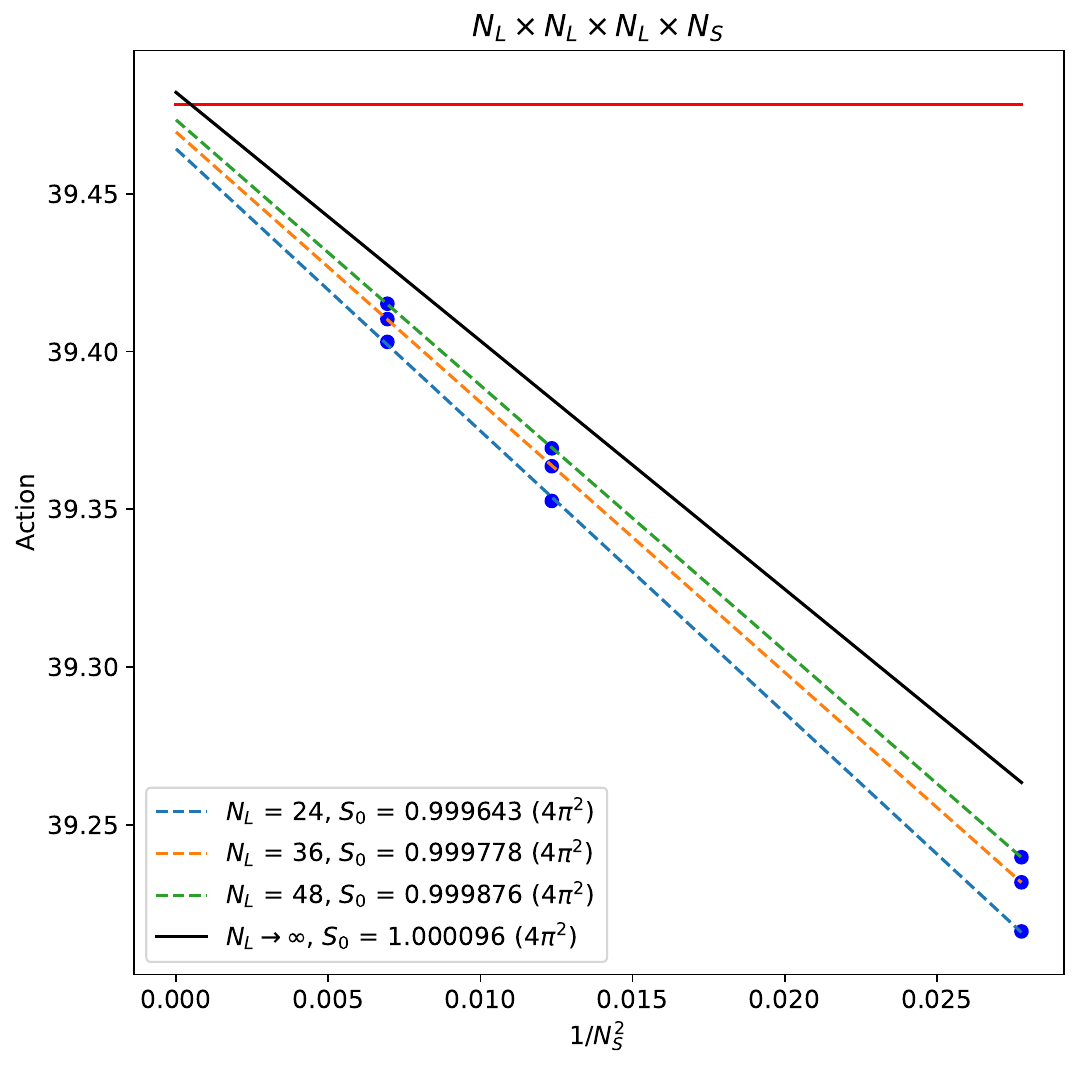}
    \caption{Fits of the final actions for lattices with the shape, $N_L\times N_L\times N_L\times N_S$. The dotted lines are linear fits with respect to $1/N_S^2$ using the data with fixed $N_L$. The solid black line is the fit of the $N_L\rightarrow \infty$ limit for each $N_S$, the $N_L\rightarrow\infty$ is calculated by a linear fit with respect to $1/N_L$. The solid red line is the value of the BPS limit in the continuum, namely $4\pi^2$.}
    \label{fig:action_fits1}
\end{figure}

\begin{figure}
    \centering
    \includegraphics[width=0.8\textwidth]{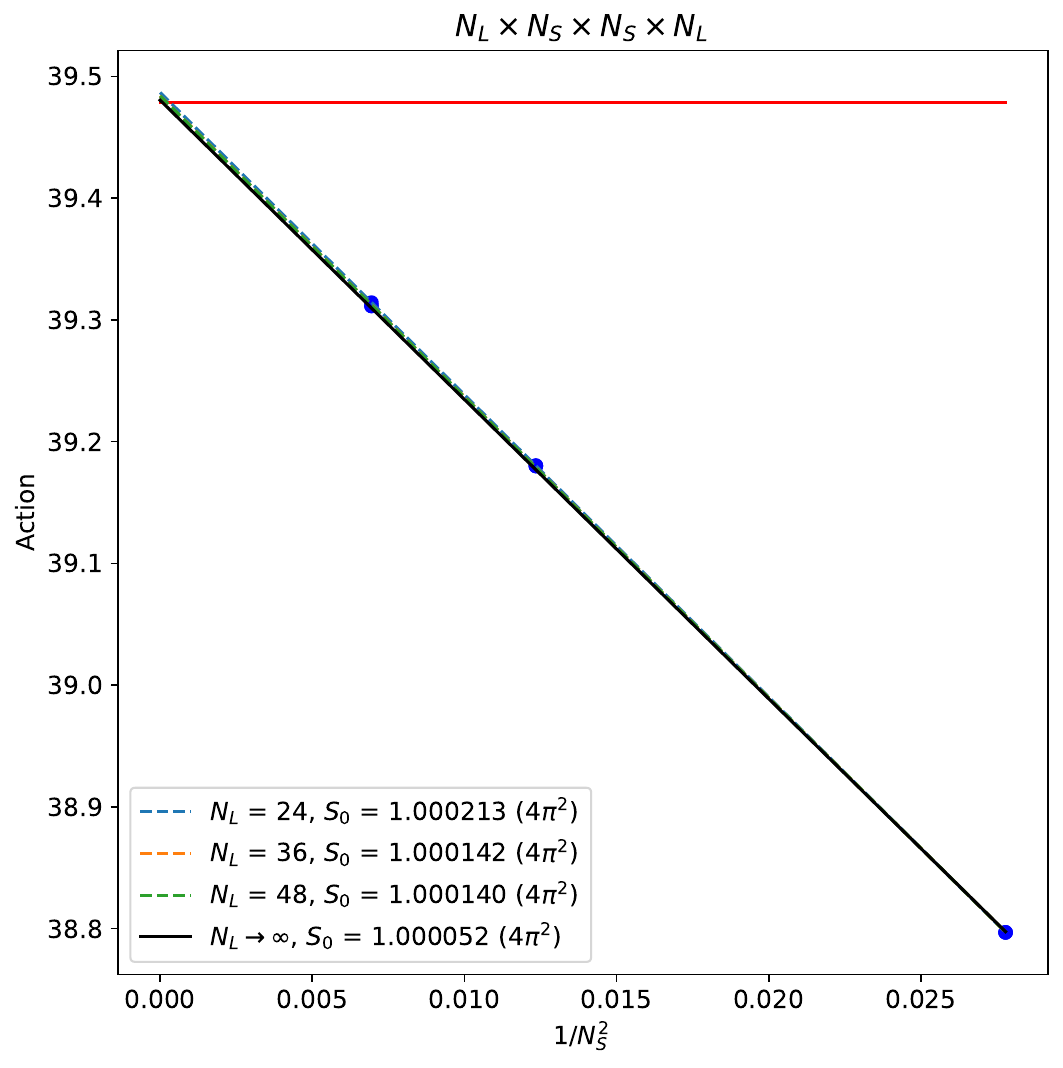}
    \caption{Fits of the final actions for lattices with the shape, $N_L\times N_S\times N_S\times N_L$. The dotted lines are linear fits with respect to $1/N_S^2$ using the data with fixed $N_L$. The solid black line is the fit of the $N_L\rightarrow \infty$ limit for each $N_S$, the $N_L\rightarrow\infty$ is calculated by a linear fit with respect to $1/N_L$. The solid red line is the value of the BPS limit in the continuum, namely $4\pi^2$.}
    \label{fig:action_fits2}
\end{figure}

\begin{figure}
    \centering
    \includegraphics[width=0.8\textwidth]{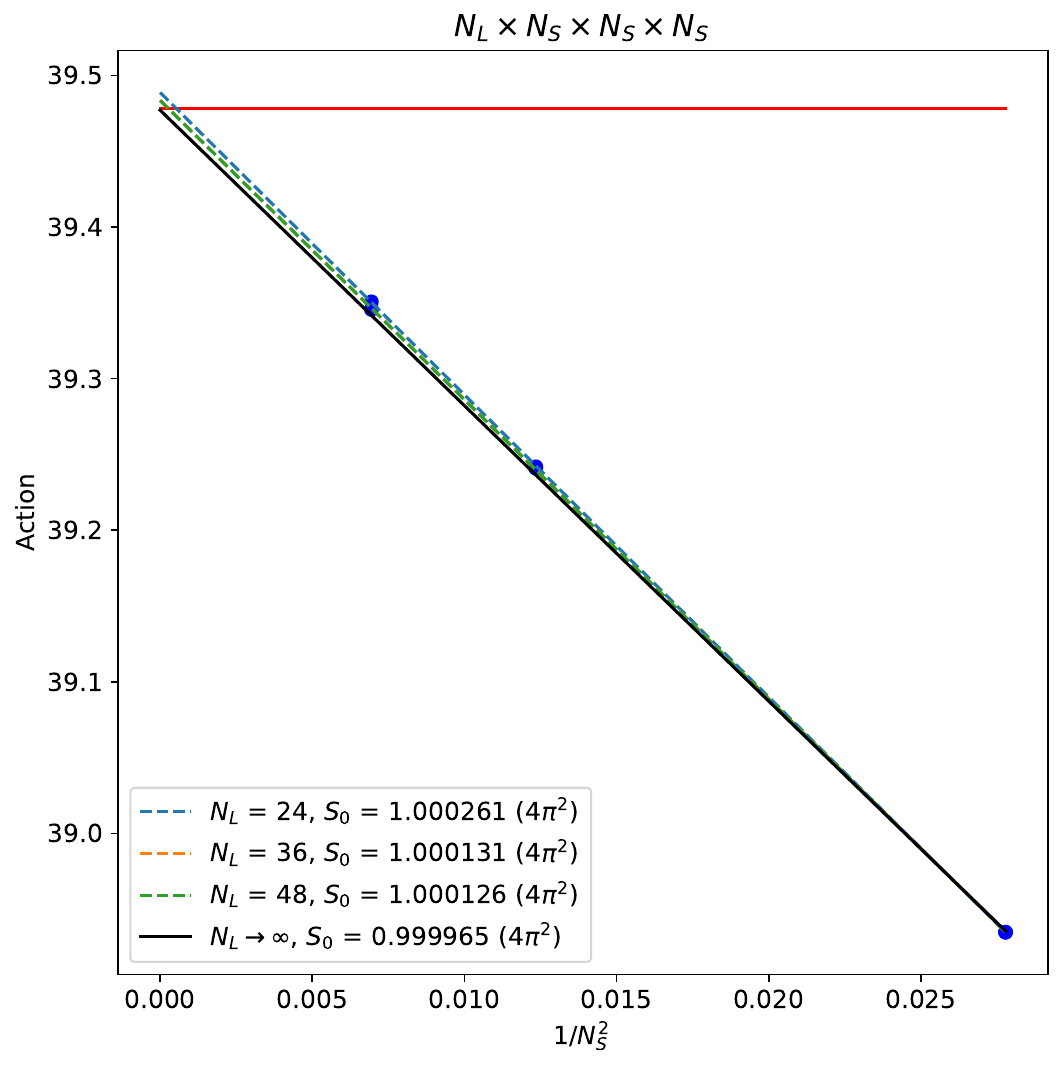}
    \caption{Fits of the final actions for lattices with the shape, $N_L\times N_S\times N_S\times N_S$. The dotted lines are linear fits with respect to $1/N_S^2$ using the data with fixed $N_L$. The solid black line is the fit of the $N_L\rightarrow \infty$ limit for each $N_S$, the $N_L\rightarrow\infty$ is calculated by a linear fit with respect to $1/N_L$. The solid red line is the value of the BPS limit in the continuum, namely $4\pi^2$.}
    \label{fig:action_fits3}
\end{figure}

Since the action depends on the square of the lattice spacing, the leading order dependence of the action on $N_S$ is
\begin{equation}
\label{eqn:ns_fit}
    S(N_S) = S_0 + \frac{c_1}{N^2_S}\, .
\end{equation}
Completing a least square error fit on the constants $S_0$ and $c_1$ for each value of $N_L$, we find that the linear approximations given by the dashed lines in Figures \ref{fig:action_fits1}, \ref{fig:action_fits2}, and \ref{fig:action_fits3}. The infinite $N_S$ limits give a much better agreement with the expected BPS bound. 

However, taking the infinite $N_S$ limit at fixed $N_L$ is questionable. Moreover, we see an slight improvement with increased $N_L$ that would be nice to include. To properly take the dependence on $N_L$ into account, we start by first taking the $N_L\rightarrow\infty$ limit. Assuming a rational dependence on $N_L$, we fit the leading order dependence,
\begin{equation}
    S(N_S,N_L) = S_0(N_S) + \frac{c_2}{N_L} \, ,
\end{equation}
for each value of $N_S$. We can then fit the resulting $S_0(N_S)$ according to \ref{eqn:ns_fit}. This fit gives the solid black lines in Figures \ref{fig:action_fits1}, \ref{fig:action_fits2}, and \ref{fig:action_fits3}. Here we see an even closer agreement between the continuum limiting value, $S_0$, and the expected saturation of the BPS bound. This evidence suggests that the discrepancy between the expected and calculated actions is a lattice effect, and, therefore, that our solutions are approximating true BPS states.

\subsection{Moduli space and antiinstantons}
\label{sec:modulispace}

One advantage of being able to generate large datasets of configurations is that it allows for statistically reliable statements regarding the moduli space of configurations. We start by making an educated guess about moduli space, then we generate 300 configurations from uniformly distributed random initial configurations. If all 300 of the configurations end up within the expected moduli space of solutions, then the "Rule of Three" \cite{ruleof3} dictates that, with 95\% confidence, the the probability for the algorithm to find a solution outside of the expected moduli space is less than 1\%. Since we are uniformly (with respect to the Haar measure) randomly sampling the space of initial configuration, this is equivalent to the statement that less than 1\% of the configuration space of the lattice will minimize out of the expected moduli space, with 95\% confidence. This gives us a statistically rigorous way of stating how confident we are that other fractional instanton solutions do not exist\footnote{The space of initial configurations is very high dimensional, so 1\% still represents a large unexplored subset of the parameter space where alternative solutions might exist. However, if other fractional instanton solutions exist, they must have the unusual property of being heavily disfavoured by this algorithm. Without any clear reason for bias in the cooling algorithm, it seems safe  to assume that we are not missing any regions of moduli space.}. 

To establish the expectation for the moduli space, consider the large Wilson loops, $W_i$. The twist in the $12$-plane means that $W_1$ is antiperiodic in $x^2$ and $W_2$ is antiperiodic in $x^1$. Similarly, the twist in the $03$-plane means that $W_0$ is antiperiodic in $x^3$ and $W_3$ is antiperiodic in $x^0$. This fact is equivalent to the statement that a shift by $N_i$ in the $i^\text{th}$ direction is equivalent to a center symmetry transformation in the $j^\text{th}$ direction, if $n_{ij}\neq 0$. Also, as we demonstrate in later sections, the $W_i$ are not identically zero for any $i$ or for any lattice shape. Hence, translating the configuration in any direction will change the gauge invariant observables, but not the action. Moreover, a translation of $2N_i$ (or a multiple thereof) is required in the $i$-direction to return to the initial configuration. This suggests that the moduli includes a 4-torus that is twice the size of the system in all directions. Hence, in the continuum, we expect the moduli space to be
\begin{equation}
    \label{eqn:modspace}
    S^1_{2L_0}\times S^1_{2L_1}\times S^1_{2L_2} \times S^1_{2L_3}\,,
\end{equation}
where the $L_i$ is the length of the $i$-direction in spacetime.
Index theorems tell us that in the continuum limit, there should only be 4 continuous directions to the moduli space, suggesting that this torus is all of moduli space. Of course, there might be other disconnected regions of moduli, which is why we employ the statistical approach introduced above.

\begin{figure}
    \centering
    \includegraphics[width=0.9\textwidth]{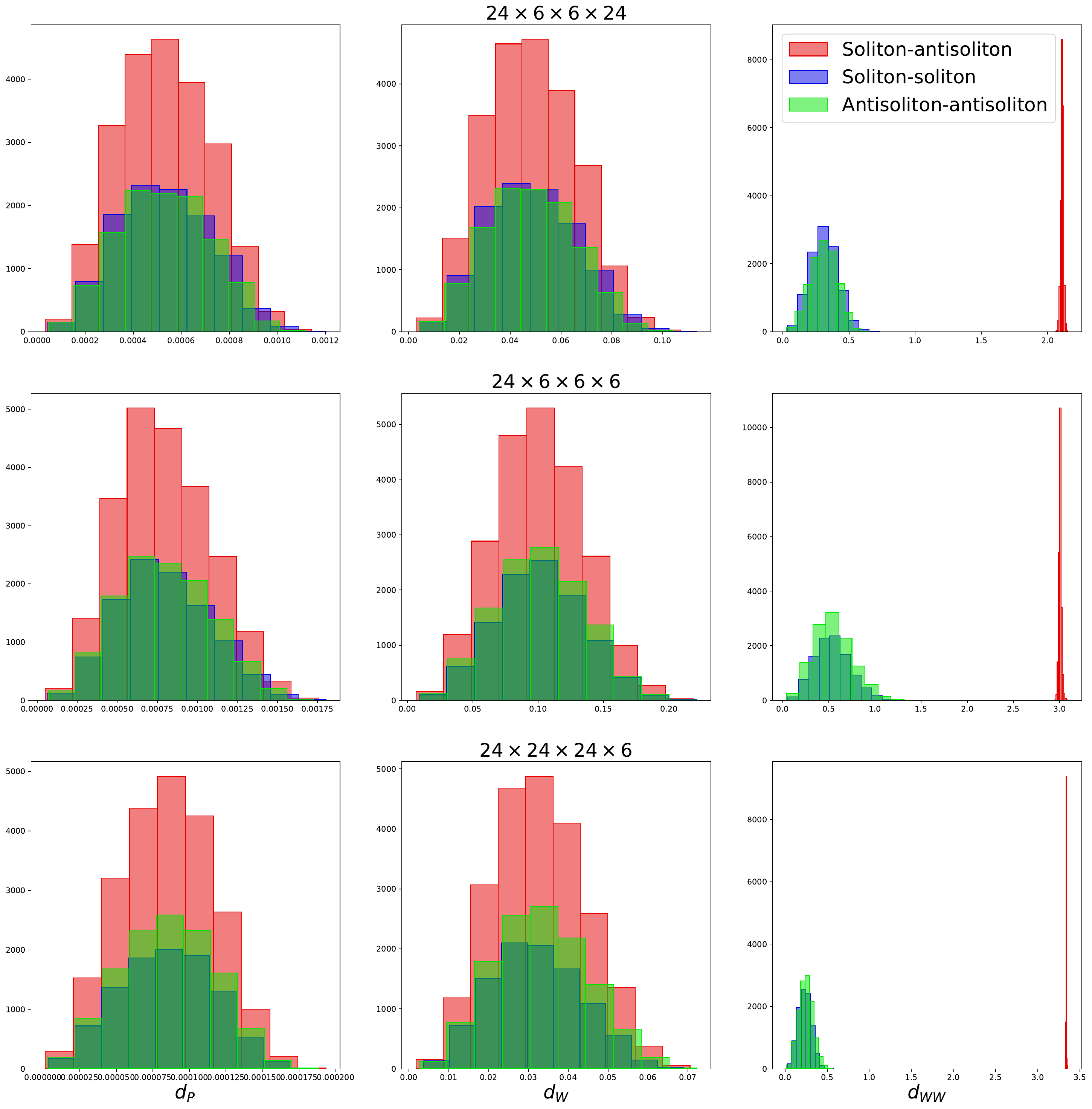}
    \caption{Comparison of the three distance metrics, $d_P$, $d_W$, and $d_{WW}$, across the three large datasets of fractional instanton configurations. The rows correspond to the datasets with lattice configurations in the shape $24\times6\times6\times24$, $24\times6\times6\times6$, and $24\times24\times24\times6$, respectively. The columns correspond to the different distance metrics. The left column shows the results for $d_P$. The central column shows results for $d_W$. The right column shows results for $d_{WW}$. Distances are measured between every possible pairing of the 300 configurations, divided based on whether they compare instantons or antiinstantons, and plotted as a histogram. The x-axis gives the distance value and the y-axis is the frequency with which that value occurs in the dataset. Notice that only the $d_{WW}$ metric captures the distinction between instantons and antiinstantons.}
    \label{fig:config_distances}
\end{figure}

One important consideration is the fact that the algorithm can also produce antiinstantons configurations. Antiinstantons have the same action as the instantons but with the opposite sign of topological charge, and they should have an identical moduli space. As mentioned above, the twists guarantee only that our configurations have half integer topological charge. Hence, fractional antiinstanton configurations in $\SU{2}$ are not ruled out by the twist, so we should expect them to show up in the dataset of configurations. This will appear like a disconnected region of moduli space in our statistical analysis (see Figure \ref{fig:config_distances}). However, since we know that these configurations will show up, we can still apply our statistical approach to rule out configurations outside of the fractional instanton and fractional antiinstanton torii. For brevity, we discuss this system as having a single moduli space including both the instanton and antiinstanton moduli spaces as disconnected regions. 

We can accurately distinguish the instantons from the antiinstantons by approximately calculating the topological charge. We do this by calculating the following gauge invariant quantity:
\begin{equation}
    \label{eqn:Qtop_discrete}
    Q_{top} = -\frac{1}{32\pi^2} \sum_{x} \varepsilon^{\mu\nu\rho\lambda} \Tr\left[\square_{\mu\nu}(x) \square_{\rho\lambda}(x)\right]
\end{equation}
where summation is implied over the indices. This is simply the naive discrete version of the continuum definition of topological charge.

\begin{figure}
    \centering
    \includegraphics[width=0.9\textwidth]{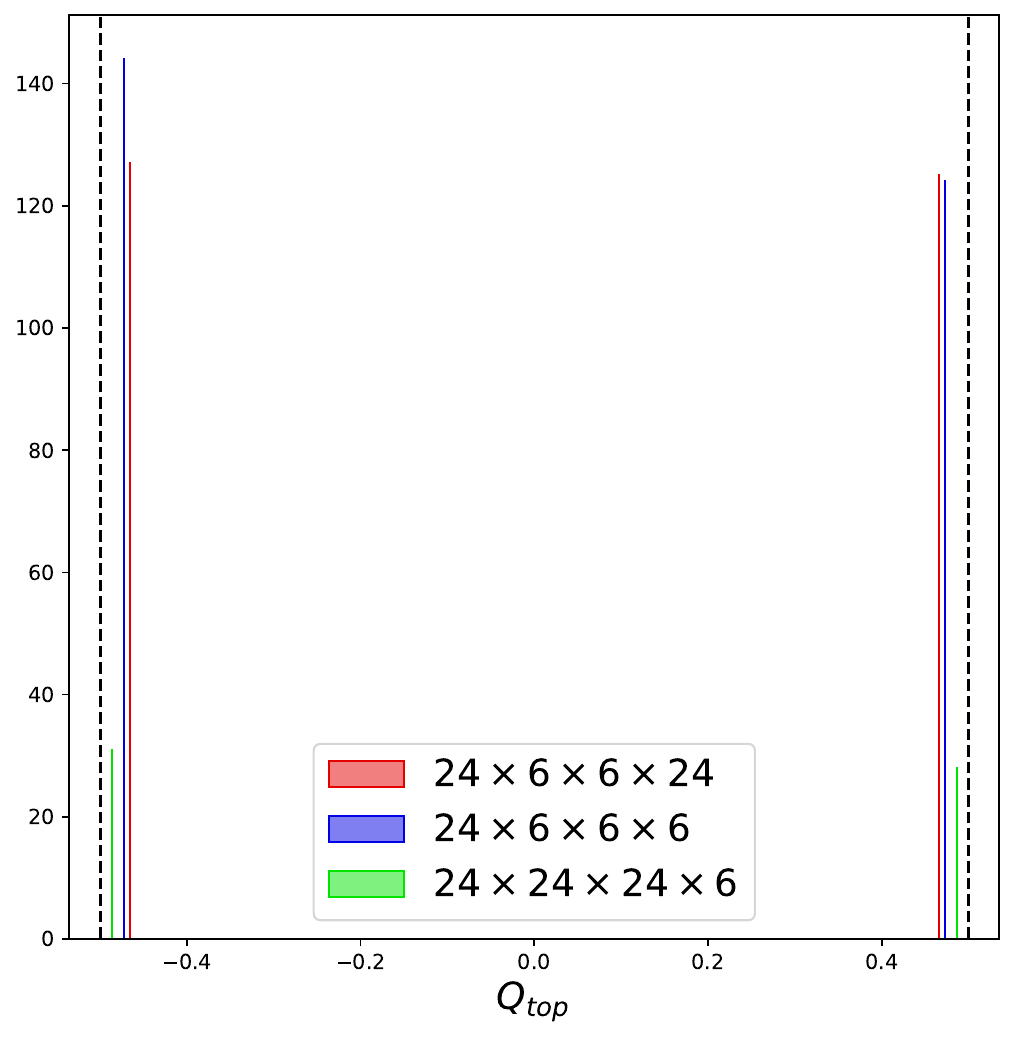}
    \caption{Histogram of the topological charges (calculated according to \ref{eqn:Qtop_discrete}) of all the final configurations in the large datasets with $n_{03}=1$. The y-axis value corresponds to the usual histogram frequency. Here the individual bins are too narrow to make out, but the relative heights of the different distributions give an indication of the spread/variance of these distributions. For example, the green distribution is shorter than the blue and red distributions because it is more spread out. The values were separated into positive and negative values before generating the histograms. The black dashed lines indicate the ideal continuum values of $\pm1/2$.}
    \label{fig:top_charges}
\end{figure}

More sophisticated methods for determining topological charge from a lattice configuration are known in the literature\footnote{See \cite{woit1985topcharge} for one of the earliest examples and \cite{Alexandrou:2017hqw} for a comparison of modern techniques.}, but this naive approach is powerful enough to distinguish the instantons from the antiinstantons. Figure \ref{fig:top_charges} demonstrates that the values are clearly clustered near the values of $\pm 1/2$.

In implementing our statistical strategy, it would be wildly impractical to compare all 300 configurations by hand, so instead we develop an automated approach. The idea is to align all the configurations as best as possible, by translating each configuration such that the peaks of the action density are at the same point and that the signs of the $W_i$ observables match at their peaks. This fixes a particular location in each connected component of moduli space. If the configurations belong to the same connected region of moduli, all of their gauge invariant quantities should agree (to within reasonable numerical error) after these translations are applied. 

To test this, we use a gauge-invariant metric that quantifies the differences between configurations. The metric involves summing the squared differences of a large number of gauge-invariant quantities. This means that the metric can only approach 0 when all these gauge invariant quantities are similar between the configurations. The physical configuration space has $9N$ degrees of freedom\footnote{Here $N = N_0N_1N_2N_3$, the total number of lattice sites. There are $4N$ links, each with 3 degrees of freedom associated with an $\SU{2}$ matrix, so the total configuration space has $12N$ degrees of freedom. However, gauge transformations are associated with $\SU{2}$ matrices at each lattice site, so the space of gauge transformations is $3N$-dimensional. Hence, the number of physical degrees of freedom is $12N-3N=9N$.}, so the metric must compare at least $9N$ gauge-invariants to properly distinguish different configurations.

To start, consider a metric defined using all possible plaquettes:
\begin{equation}
    d^2_{P}\left(U,U'\right) = \frac{1}{N} \sum_{x} \sum_{0\leq \mu < \nu \leq 3} \left(\Tr \square_{\mu\nu}(x) - \Tr\square'_{\mu\nu}(x)\right)^2
\end{equation}
where we included a factor of $\frac{1}{N}$ as a practical measure to keep the resulting numbers easy to work with. This is a useful starting point for our metric, but it only compares $6N$ gauge-invariants, hence there are distinct configurations that are not distinguished by this metric. For example, consider Figure \ref{fig:config_distances}: the left column shows the evaluation of the metric, $d_P$, between 300 configurations including instanton and antiinstanton solutions. It is clear that $d_P$ is unable to distinguish instantons from antiinstantons, even though we know they can be distinguished by gauge-invariants such as the topological charge. We must compare additional gauge-invariant quantities to make a proper metric.

Since the plaquette observables describe the local physics around a given site, it is natural to look for non-local observables to round out the metric. A first choice might be to include the holonomies around each direction:
\begin{equation}
    d^2_{W}\left(U,U'\right) = \frac{1}{N}\sum_{x}\sum_{0\leq\mu\leq 3} \left(\Tr W_\mu(x) - \Tr W'_\mu(x)\right)^2 ~.
\end{equation}
Unfortunately, since the holonomy in the $i$-direction is independent of the value of $x^i$ where it starts, there are fewer than $N$ quantities\footnote{The total number of independent gauge-invariants in $d_W$ is $N_W = N_0N_1N_2 + N_0N_1N_3 + N_0N_2N_3 + N_1N_2N_3$. If $N_i>4$ for all directions (as it is for all the configurations we consider here), then 
\begin{equation}
    \begin{split}
        4N_W = & N_0N_1N_24 + N_0N_14N_3 + N_04N_2N_3 + 4N_1N_2N_3\\
        < & N_0N_1N_2N_3 + N_0N_1N_2N_3 + N_0N_1N_2N_3 + N_0N_1N_2N_3\\
        = & 4N.
    \end{split}
\end{equation}
Hence, $N_W < N$.} compared in $d_W$. To remedy this lack of information, we can also consider products of holonomies in orthogonal directions:
\begin{equation}
    d^2_{WW}\left(U,U'\right) = \frac{1}{N}\sum_{x}\sum_{0\leq\mu<\nu\leq 3}  \left(\Tr\left[W_\mu(x)W_\nu(x)\right] - \Tr\left[W'_\mu(x)W'_\nu(x)\right]\right)^2 ~.
\end{equation}
This introduces another $6N$ gauge-invariant quantities. Importantly, it distinguishes configurations that could not be distinguished with just $d_P$ and $d_W$, namely it distinguishes the instantons from the antiinstantons. 

Using the above quantities, we can define a suitable metric via
\begin{equation}
    d^2 = d_P^2 + d_W^2 + d_{WW}^2 ~.
\end{equation}
However, it is instructive to plot each part of this metric separately. 

It is important to note that we have not proven that this collection of gauge-invariants is strong enough to distinguish every distinct configuration. It compares more than $12N$ gauge-invariant quantities, which implies the values of these gauge-invariants cannot be completely independent from each other (since they are measuring a $9N$-dimensional configuration space). It is possible, in principle, that the true number of independent quantities measured by $d$ is actually less than the necessary $9N$; however, this seems unlikely as this would require a deep correlation between these quantities, which is unknown to the author. The fact that $d_{WW}$ contains information that is not contained in $d_P$ or $d_W$ (as evidenced by its ability to distinguish instantons from antiinstantons) suggests there is a large degree of independence between the considered quantities. 


The results of comparing all 300 configurations for the $24\times24\times24\times6$, $24\times6\times6\times24$, and $24\times6\times6\times6$ lattice sizes are given in Figure \ref{fig:config_distances}. This suggests that there are no populations of solutions significantly differing from the proposed instanton and antiinstanton. As stated above, this does not rule out alternative solutions, but severely constrains them. The volume of parameter space that minimizes down to a different solution is less than 1\% of the total volume of parameter space at the 95\% confidence level. Moreover, at 99.99994\% confidence level, the volume is constrained to be less than 3.33\% of the total space. This suggests that the moduli space for instantons is the torus described in \ref{eqn:modspace}, and the same for the antiinstantons.

Given that this holds in the three extremal cases presented above, it is natural to assume that this also holds for all sizes of torii considered in this paper. \footnote{There may be an issue with this, however, near the case of the self-dual torus (i.e. $\Delta\approx0$). This case was considered recently in a calculation of the gaugino condensate in SYM \cite{Anber:2022qsz}, where it was found that this double circumference torus gives a result that is a factor of 2 greater than the infinite volume result. In Section \ref{sec:interpolation}, we give some preliminary evidence that the same moduli space does agree with \ref{eqn:modspace}.}

\section{Center vortices on $\RT$}
\label{sec:center_vortices}
In this section, we discuss the properties of the solutions with small $N_1$ and $N_2$ and large $N_0$ and $N_3$. This is our best approximation to the spacetime $\R^2\times\T^2_*$, where the confinement mechanism is understood in terms of BPS center vortices appearing in \cite{Tanizaki:2022ngt}. The use of center vortices in \cite{Tanizaki:2022ngt} was based on numerical evidence from \cite{GONZALEZARROYO1998273} which we are expanding on and strengthening in this work. The key feature needed for the semiclassical argument of confinement in $\RT$ is that the center vortices are localized in $\R^2$ (in our case in the 03-plane), link with Wilson loops embedded in $\R^2$, and are surrounded by the vacua shown in \ref{eqn:vacua}. We confirm all of these properties below.

\begin{figure}
    \centering
    \includegraphics[width=0.9\textwidth]{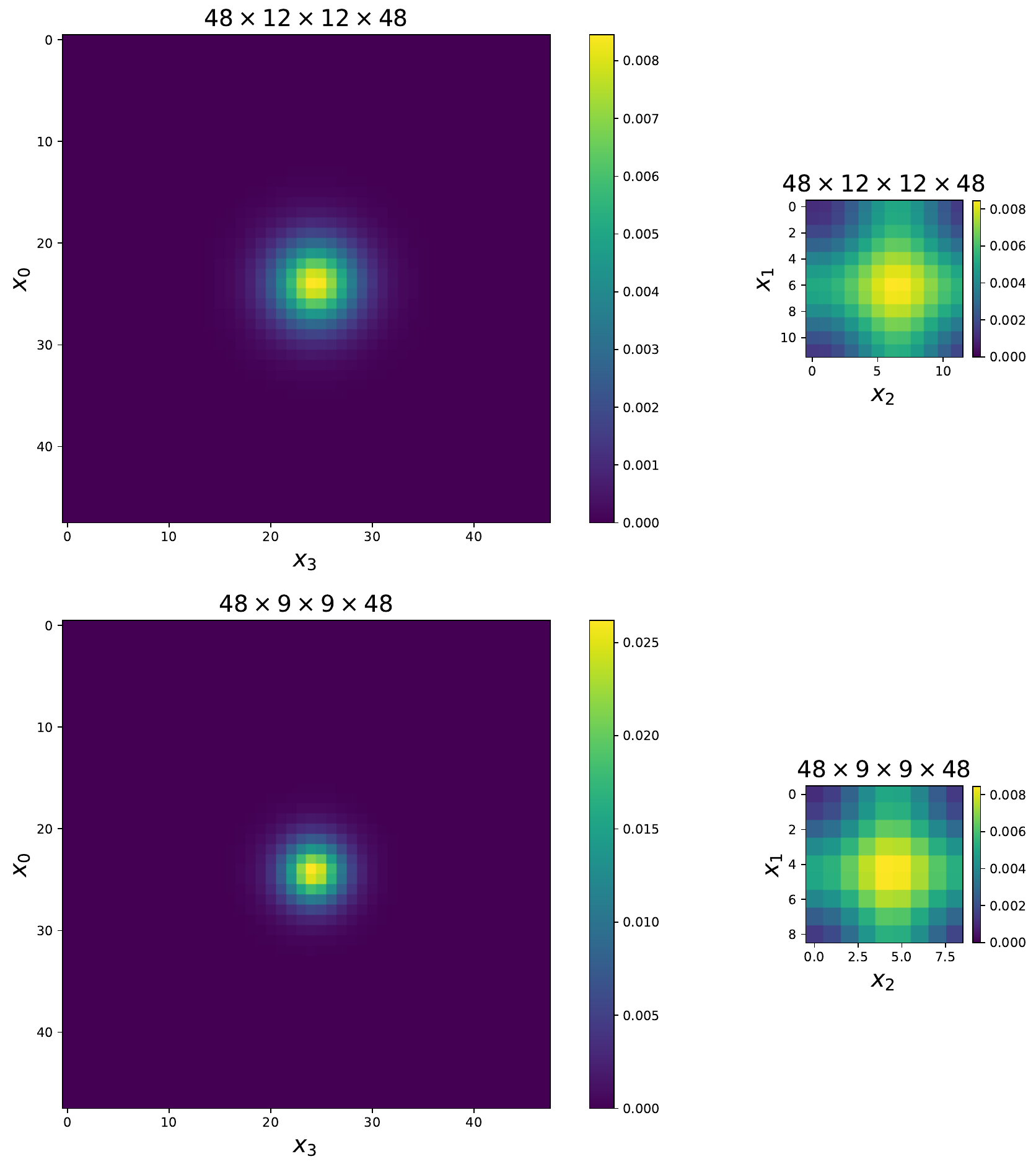}
    \caption{Plots of the action density across the $03$- and $12$-planes that cut through the point of maximal action density. Both the $48\times12\times12\times48$ and $48\times9\times9\times48$ configurations are shown here to highlight how the size of the localized instanton changes with the size of the small directions.}
    \label{fig:SDen_centervortex}
\end{figure}

With regards to localization, consider Figure \ref{fig:SDen_centervortex} which shows the action density of the resulting configurations at various lattice shapes. Two things are clear from these plots: the solutions are localized in the 03-plane and the size of the vortex depends on the size of the small 12-plane. This matches the expectation for the infinite volume limit.

\begin{figure}
    \centering
    \includegraphics[width=0.7\textwidth]{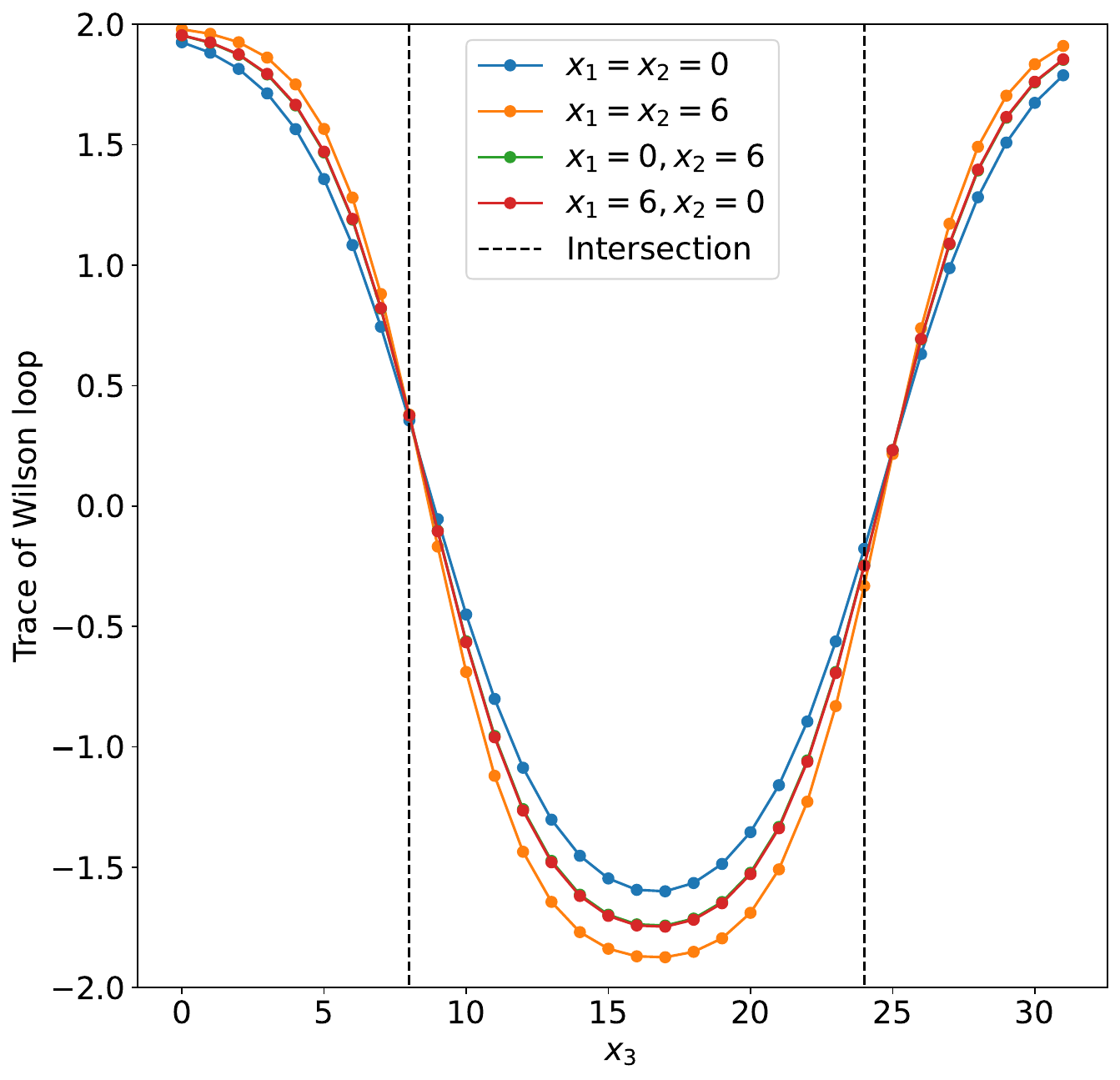}
    \caption{This plot shows the trace of a rectangular Wilson loop embedded in the 03-plane of a $48\times12\times12\times48$ solution as the Wilson loop is moved across the solution in the 3-direction. The Wilson loop is 46 by 16, and the different lines show Wilson loops embedded at different point in the $12$-plane. The dotted black lines represent the points where the edges of the Wilson loop cross the peak of the center vortex. The x-axis gives the value of $x^3$ at the edge with the lowest value of $x^3$.}
    \label{fig:linking_centervortex}
\end{figure}

A peculiar feature of these solutions is the peak in the action density within the 12-plane. This raises the concern that the solutions might localize in the 12-plane, which would imply that they cannot be center vortices, since center vortices must be codim-2. However, we can confirm that our solutions are center vortices by demonstrating that our solutions properly link with Wilson loops embedded anywhere in the 12-plane. In the full continuum quantum theory, linking between a Wilson loop operator, $\hat{W}(C)$, defined on a contour $C$ and a center vortex, $\hat{V}(\Sigma)$, defined on a surface $\Sigma$ is defined by the relation:
\begin{equation}
    \hat{W}(C) \hat{V}(\Sigma) = \exp{i \frac{2\pi}{N}\text{Linking}(C,\Sigma)} \hat{V}(\Sigma) \hat{W}(C)
\end{equation}
where $\text{Linking}(C,\Sigma)$ is the linking number between $C$ and $\Sigma$. In the case of semiclassical $\SU{2}$, this relation simplifies to the statement that a Wilson loop that links with the center vortex must have opposite sign to a Wilson loop that does not link with it. In our solutions, a Wilson loop in the 03-plane is linked by the center vortex if the cross section of the center vortex in the 03-plane is contained within the Wilson loop. Hence to test the linking property of our solutions, it is sufficient to measure the values of large Wilson loops embedded parallel to the 03-plane in various positions. These values is shown in Figure \ref{fig:linking_centervortex} for a rectangular Wilson loop that is translated across the 03-plane at various points in the 12-plane. Notice that the change in sign that indicates proper linking with the center vortex occurs at all point in the 12-plane, even when the action density is minimal. This indicates that the linking is robust and the solutions are truly codim-2 center vortices.

\begin{figure}
    \centering
    \includegraphics[width=0.7\textwidth]{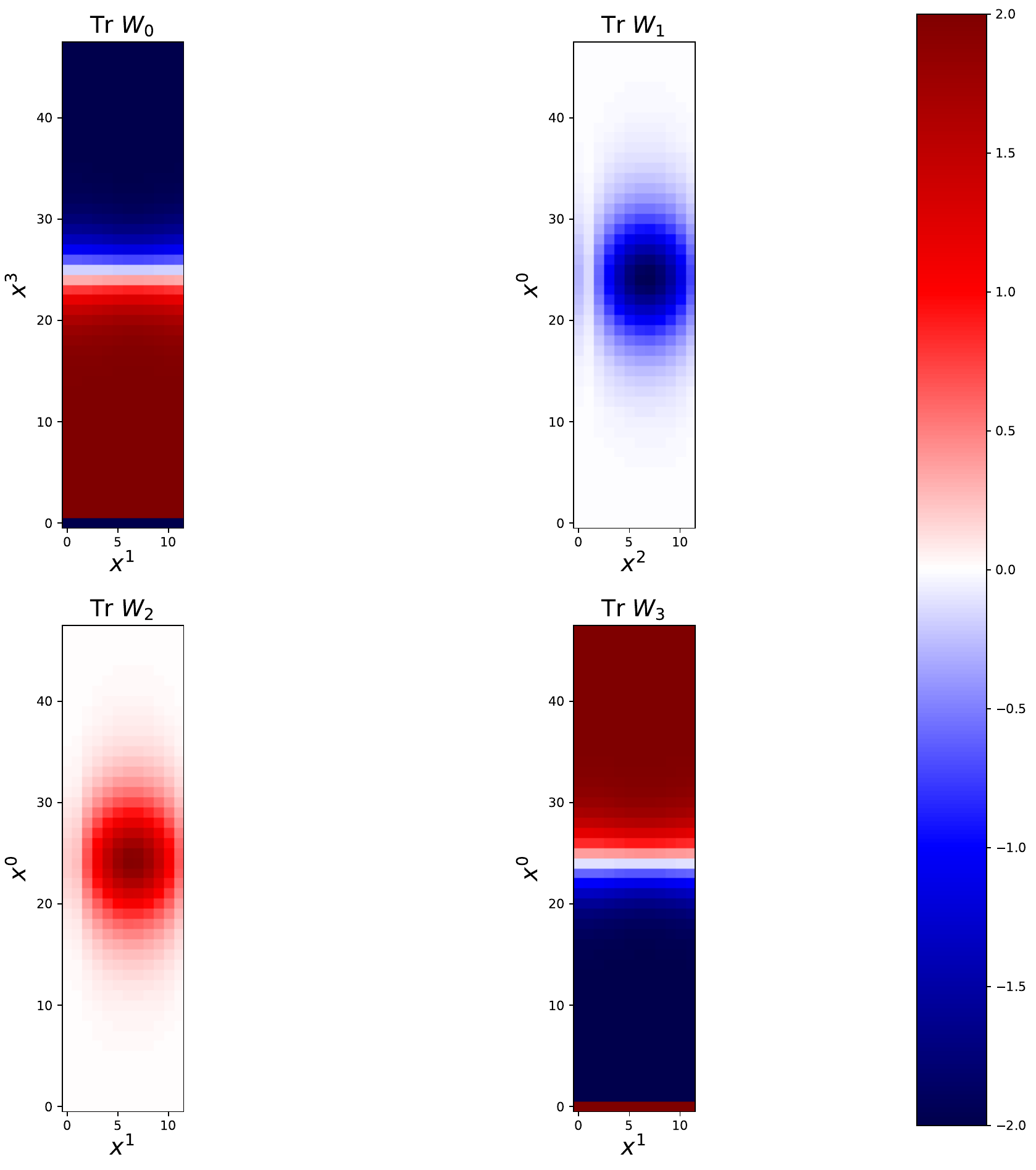}
    \caption{Traces of the holonomies of the $48\times12\times12\times48$ lattice configuration. Holonomies are plotted along planes that intersect the point of maximal action density. Discontinuities are due to the fact that the holonomies are antiperiodic across twists, which are always contained in the plaquettes at the origin.}
    \label{fig:holonomy_centervortex}
\end{figure}

To show that the vacua surrounding the vortices have the correct form, we simply need to check the values of the holonomies in all four directions. The results are collected in Figure \ref{fig:holonomy_centervortex}. It is clear that away from the center of the vortex, $\Tr W_1$ and $\Tr W_2$ vanish as expected. $\frac{1}{2}\Tr W_0$ interpolates between +1 at $x^3=1$ to -1 at $x^3=N_3$, and similarly $\frac{1}{2}\Tr W_3$ interpolates between -1 at $x^0=1$ to +1 at $x^3=N_0$. We also verify that $\Tr W_1W_2$ tends to zero asymptotically. Therefore, when the center vortex is centered in the 03-plane, it splits the plane into quadrants, each of which contains one of the four possible solutions from \ref{eqn:vacua}. 

These observations confirm the identity of the half-BPS instantons as center vortices. Combined with the statistical bounds on other half-BPS states from Section \ref{sec:modulispace}, this provides solid numerical evidence to back the semiclassical calculations of confinement in \cite{Tanizaki:2022ngt}.

\section{Monopole instantons on $\RS$}
\label{sec:monopoles}
In this section we focus on solutions when the lattice has small $N_3$ and large $N_0$, $N_1$, and $N_2$. This approximates the case of $\RS$, where semiclassics relies on monopole-instantons. The key features of monopole instantons are that they are 1-dimensional and localized in all directions except the small circle direction, that the asymptotic vacua surrounding them will abelianize the theory to a $U(1)$ subgroup, and that with respect to this $U(1)$ they have a well-defined magnetic charge quantized to $4\pi$.

\begin{figure}
    \centering
    \includegraphics[width=0.9\textwidth]{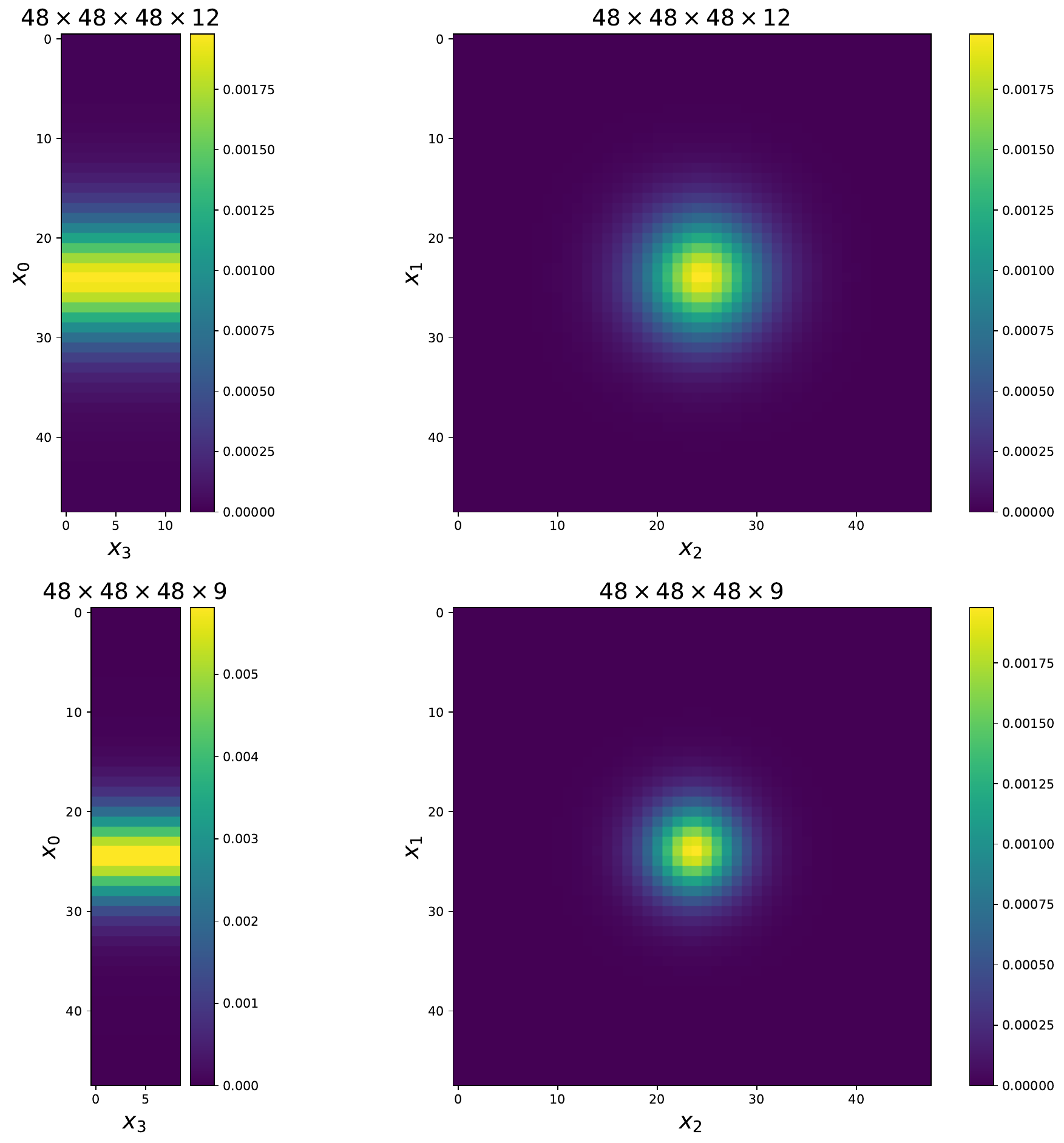}
    \caption{Plots of the action density across the $03$- and $12$-planes that cut through the point of maximal action density. Both the $48\times48\times48\times12$ and $48\times48\times48\times9$ configurations are shown here to highlight how the size of the localized instanton changes with the size of the small directions.}
    \label{fig:SDen_monopole}
\end{figure}

The localization of our classical solution is apparent in the plot of the action density presented in Figure \ref{fig:SDen_monopole}. Notice that in the $12$-plane the excitation is localized to a small disk and in the $03$-plane the excitation is contained in a thin band along the 3-direction. The fact that this solution has constant action density along the 3-direction is a good indication that these are the expected 1-dimensional monopole-instantons. 

\begin{figure}
    \centering
    \includegraphics[width=0.7\textwidth]{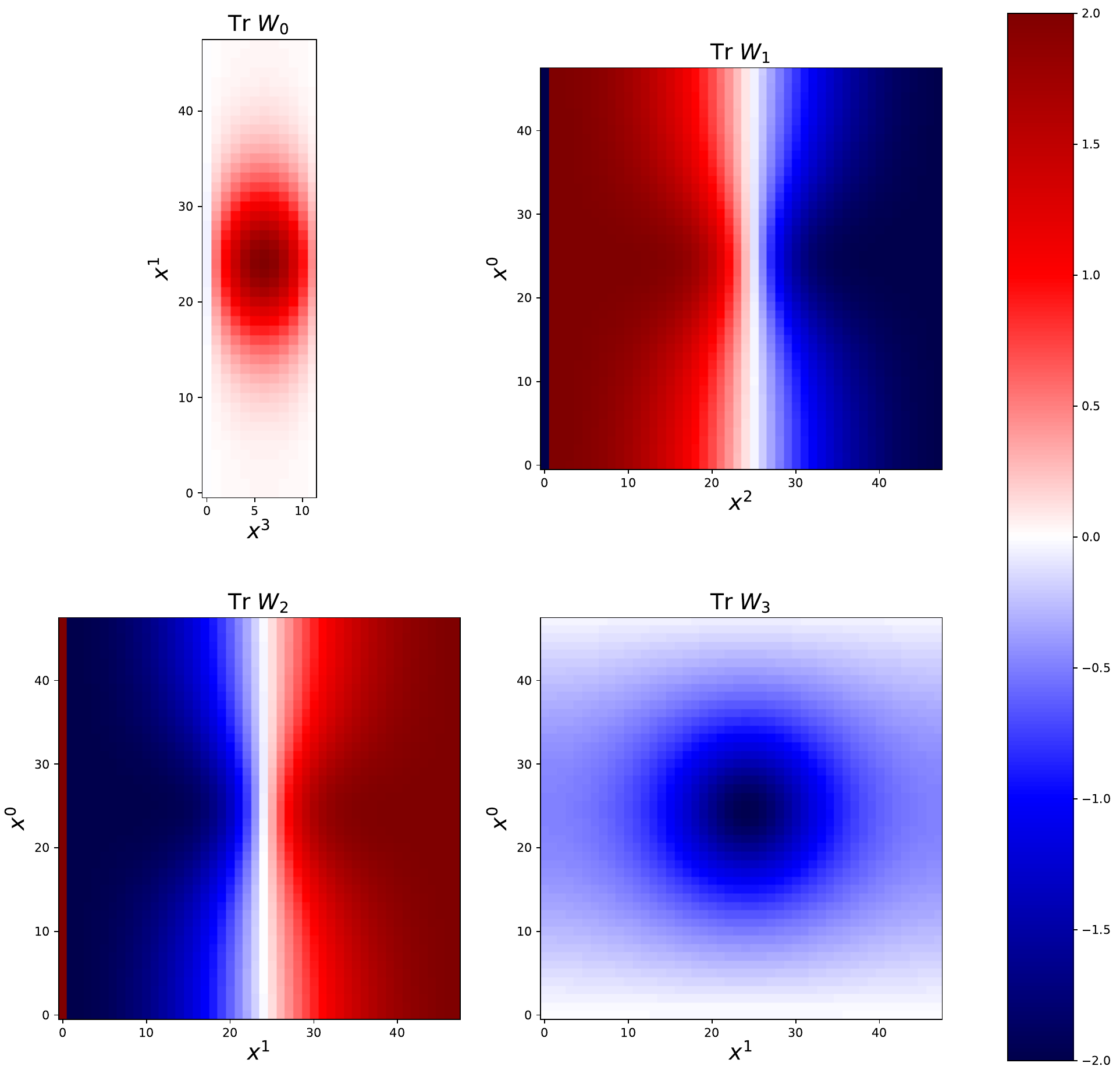}
    \caption{Traces of the holonomies of the $48\times48\times48\times12$ lattice configuration. Holonomies are plotted along planes that intersect the point of maximal action density. Discontinuities are due to the fact that the holonomies are antiperiodic across twists, which are always contained in the plaquettes at the origin.}
    \label{fig:holonomy_monopole}
\end{figure}

The external vacua are characterized by the values of the holonomies shown in Figure \ref{fig:holonomy_monopole}. In contrast to the center vortex case, where the asymptotic vacua agreed with \ref{eqn:vacua}, the vacua around the monopole-like solutions follow
\begin{equation}
\begin{split}
    \frac{1}{2} \Tr W_1 = & \pm 1 \\
    \frac{1}{2} \Tr W_2 = & \pm 1 \\
    \frac{1}{2} \Tr W_3 = & 0 \\
    \frac{1}{2} \Tr W_0 = & 0 \\
\end{split}
\end{equation}
with $\Tr W_0W_3 = 0$. By symmetry, these are the vacua in the case where there is a twist in the $03$-plane and not in the $12$-plane. 

Since we are approximating $\RS$, $W_3$ controls whether the theory abelianizes. The asymptotic vacuum has $\Tr W_3=0$, meaning the theory outside the localized excitation abelianizes from $\SU{2}$ to $U(1)$. This is desirable for defining a magnetic field and characterizing the instanton as a monopole.

The specific unbroken $U(1)$ aligns with the direction of $W_3$ in colour-space. Given that $\square_{\mu\nu}(x)\sim e^{ia^2 \vec{F}_{\mu\nu}(x)\cdot \vec{\sigma}}$ and that $\Tr W_3 = 0$ implies $W_3 = i\hat{n}\cdot \vec{\sigma}$ for some unit vector $\hat{n}\in \R^3$, we can calculate the approximate magnetic field as
\begin{equation}
    \frac{1}{2} \Tr W_3(x)\square_{ij}(x) \sim a^2 \hat{n}\cdot F_{ij} \, .
\end{equation}
Here $i,j = 0,1,2$. The continuum limit of this quantity is the magnetic flux passing through the plaquette $\square_{ij}(x)$. Notice that this is a gauge invariant because the holonomy and the plaquette both start and end at the same point, $x$. 

\begin{figure}
    \centering
    \includegraphics[width=0.9\textwidth]{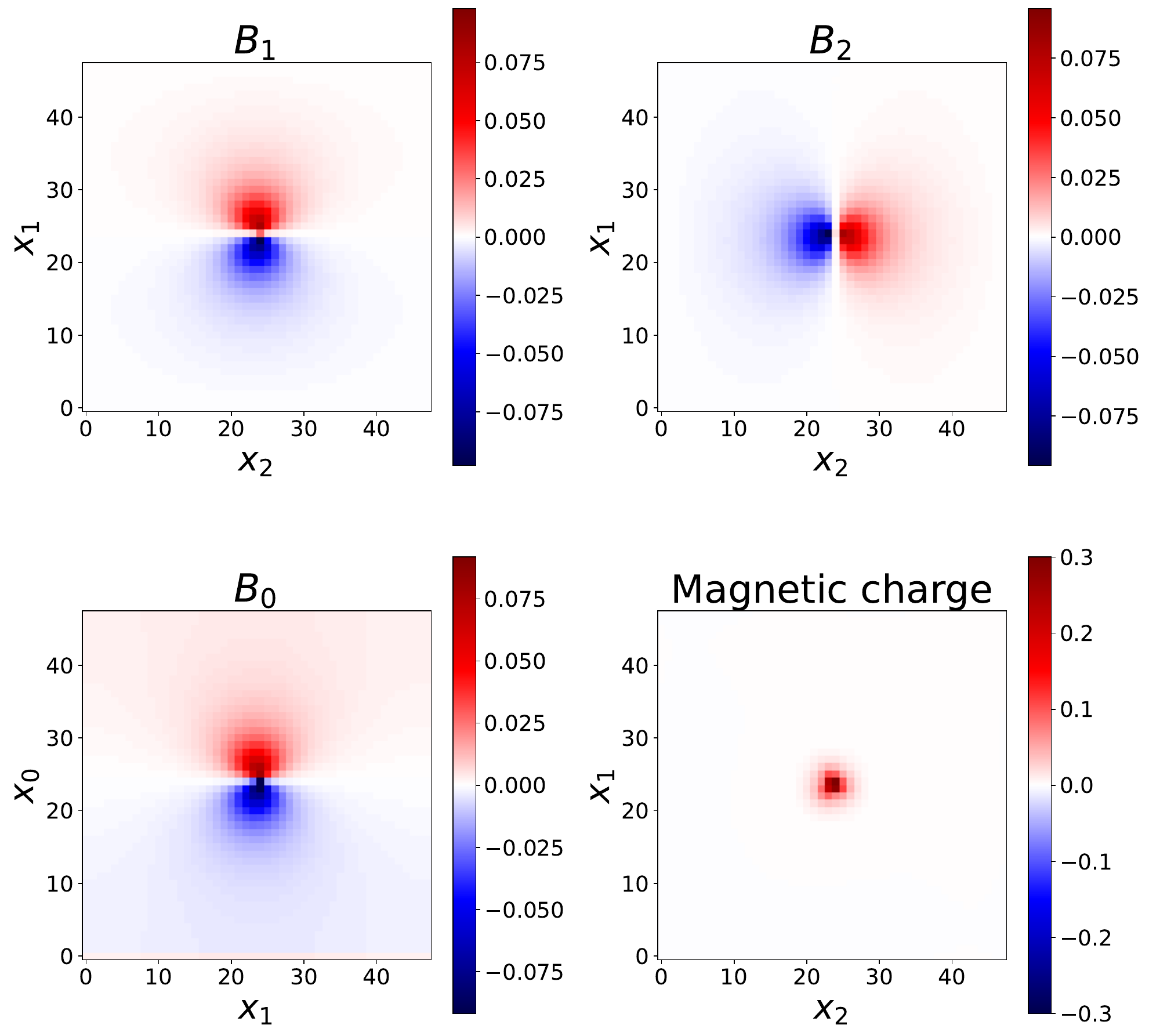}
    \caption{Plotting the magnetic flux of the $48\time48\time48\times6$ configuration calculated according to Equation \ref{eqn:Bdef}. The top left plot is the 1-component of the magnetic field, $B_1$, (calculated via the flux through the $02$-plaquettes) plotted in the $12$-plane. The top right plot is the 2-component of the magnetic field, $B_2$, plotted in the $12$-plane. The bottom left plot is the 0-component of the magnetic field, $B_0$, plotted in the $01$-plane. While $B_1$ and $B_2$ die off away from the central magnetic charge, it is clear that the $B_0$ field does not vanish as it approaches $x^0 = 1$ and $x^0=48$. The bottom left plot shows the magnetic charge distribution calculated via Gauss' Law.}
    \label{fig:Bfield}
\end{figure}

However, within finite distance of the instanton, $\Tr W_3$ will not exactly vanish giving instead $W_3 = \cos{\theta} + i\sin{\theta} \hat{n}\cdot \vec{\sigma}$ for some $\theta \in [0,2\pi)$. We want to cancel the dependence on $\theta$, so we define our approximate magnetic flux\footnote{While this quantity is technically the magnetic flux through the plaquette, $\square_{ij}(x)$, it is also directly proportional to the magnetic field near $x$. We will not be overly careful in making the distinction between the flux through a small plaquette and the field.} as
\begin{equation}
\label{eqn:Bdef}
    \varepsilon_{ijk} B_k(x) = \frac{1}{2} \left(\frac{\Tr W_3(x)\square_{ij}(x) - \Tr W_3(x)}{\sqrt{1-\left(\frac{1}{2}\Tr W_3(x)\right)^2}}\right)\, .
\end{equation}
The values of this gauge invariant quantity are plotted in Figure \ref{fig:Bfield}. It is clear that close to the center of the configuration, the magnetic field lines are pointing away from the localized instanton, consistent with a postive magnetic charge. The magnetic fields die off to zero in the $1$- and $2$-directions, but do not die off along the $0$-direction. This boundary behavior is consistent with an image charge picture where the instanton is surrounded by a lattice of image magnetic charges, with all like charges along the $1$- and $2$- directions, but alternating charges along the $0$-direction. The magnetic fields behave differently in the $0$-direction because they are antiperiodic in $x^0$, whereas they are periodic in $x^1$ and $x^2$. This is a direct consequence of $W_3$ being antiperiodic in $x^0$ but periodic in $x^1$ and $x^2$. In fact, by translating the instanton by $N_0$ sites in the $0$-direction, the sign of the magnetic charge is flipped. 

The continuum $\SU{2}$ theory on $\RS$ has two distinct types of monopoles, called BPS and KK. They have the same topological charges but opposite magnetic charges and are exchanged by a center symmetry transformation. Both types are necessary for the magnetic bion mechanism for confinement \cite{Unsal2009bions}. In our lattice configurations, it is clear that the translation by $N_0$ sites in the $0$-direction, which is a center symmetry transformation in $x_3$ and flips the sign of the magnetic fields, exchanges the BPS and KK monopoles. Hence, on the lattice, the two flavours of monopoles are related by translations and therefore are part of the same connected region of moduli space.

Using Gauss' Law, we can measure the magnetic charge contained within each cube of our lattice configuration. The results are presented in the lower right plot in Figure \ref{fig:Bfield}. Again this is consistent with a small blob of positive magnetic charge coincident with the instanton. The total amount of charge (and simultaneously the total magnetic flux) is consistent with $4\pi$ (see Table \ref{tab:magneticcharge} for how close we get at various lattice sizes). This is the expectation for a single quantum of magnetic charge in the continuum theory, hence our solutions are consistent with continuum monopole-instantons. 

\begin{table}
\centering
\begin{tabular}{|c|ccc|}
     \hline
     \diagbox{$N_S$}{$N_L$} & 24 & 36 & 48 \\
     \hline
     6 & $12.5662 \substack{+0.0008\\-0.0011}$ & 12.5662 & 12.5664 \\
     9 & $12.5532 \substack{+0.0018\\-0.0031}$ & 12.5658 & 12.5664 \\
     12 & $12.4885 \substack{+0.0031\\-0.0099}$ & 12.5597 & 12.5661 \\
     \hline
\end{tabular}
\caption{Values of the total magnetic flux for various sizes of lattice. Values with errors are averaged across a dataset of multiple runs at the same lattice size. Errors are determined by the maximum and minimum values across the dataset}
\label{tab:magneticcharge}
\end{table}

One key detail of these solutions is that the flux escapes out through the $12$-planes at $x^0=1$ and $x^0=N_0$. This suggests that the instantons are interpolating between vacua with $\Tr W_3 = 0$ and a magnetic flux of $\pm 2\pi$ though the $12$-plane. These are exactly the vacua originating in \cite{Perez:2013aa} and \cite{Unsal:2020yeh} and considered in \cite{Poppitz2023} as the proper semiclassical vacua for dYM on $\RTS$ with large $\T^2$. This is further evidence that these flux vacua are the true vacua after stabilizing $\Tr W_3 = 0$ and that these configurations are useful for semiclassics in center-stabilized YM on $\RS$.

Our solutions are useful for validating qualitative features of these monopole-instantons in center-stabilized YM, though we expect that the additional stabilizing potential will impose a compressive force that decreases the volume of the instantons. It would be illuminating to study these minimal solutions in the presence of a deformation potential, but this would require a significant modification of the cooling algorithm and thus is left to a future study.

On the other hand, in the pure (i.e. unstabilized) YM on $\RTS$ (as considered in Section \ref{sec:true_vacua}), the vacuum obeys $\frac{1}{2} \Tr W_3 = \pm 1$. Here $W_3$ is proportional to the identity; therefore, in perturbation theory around this vacuum, the theory does not abelianize. If the theory does not abelianize, then it must be strongly coupled in the IR and it is hopeless to start a controlled semiclassical calculation. This limits the usefulness of the monopole-instanton solutions for studying pure YM on $\RTS$.

Interestingly, having the twist in the $12$-plane causes $\frac{1}{2} \Tr W_3 = \pm 1$ classically, whereas in the continuum theory on $\RS$, we have $\frac{1}{2} \Tr W_3 = \pm 1$ due to the GPY potential \cite{GPY1981}, which is a purely quantum effect. In both cases, this suggests a spontaneous breaking of the center symmetry that cannot be restored by semiclassical effects. This is consistent with the high temperature deconfined phase of YM.

\section{Interpolating between the different regimes on $\RTS$}
\label{sec:interpolation}
This section is devoted to understanding how the fractional instantons on $\RTS$ interpolate between center vortices in the large $S^1$ limit and monopole-instantons in the large $\T^2_*$ limit. The investigation focuses on comparing minimal configurations on different sizes of lattice, specifically $N_0\times N_T\times N_T\times N_S$ where $N_0$ is kept large and one of $N_T$ or $N_S$ is always kept small while the other is set to intermediate values. Essentially, we test how the instantons change as we go from the $\RS$ limit to the femtouniverse limit (with $N_T$ and $N_S$ both small) to the $\RT$ limit. We find it convenient for this discussion to introduce a dimensionless parameter measuring the asymmetry between the $03$- and $12$-planes\footnote{Be warned that this definition of $\Delta$ differs slightly from other definitions due to the way we have chosen to label our indices.}:
\begin{equation}
\label{eqn:Delta}
    \Delta = \frac{N_1N_2 - N_0N_3}{\sqrt{N_0N_1N_2N_3}} \, .
\end{equation}

\begin{figure}
    \centering
    \includegraphics[width=\textwidth]{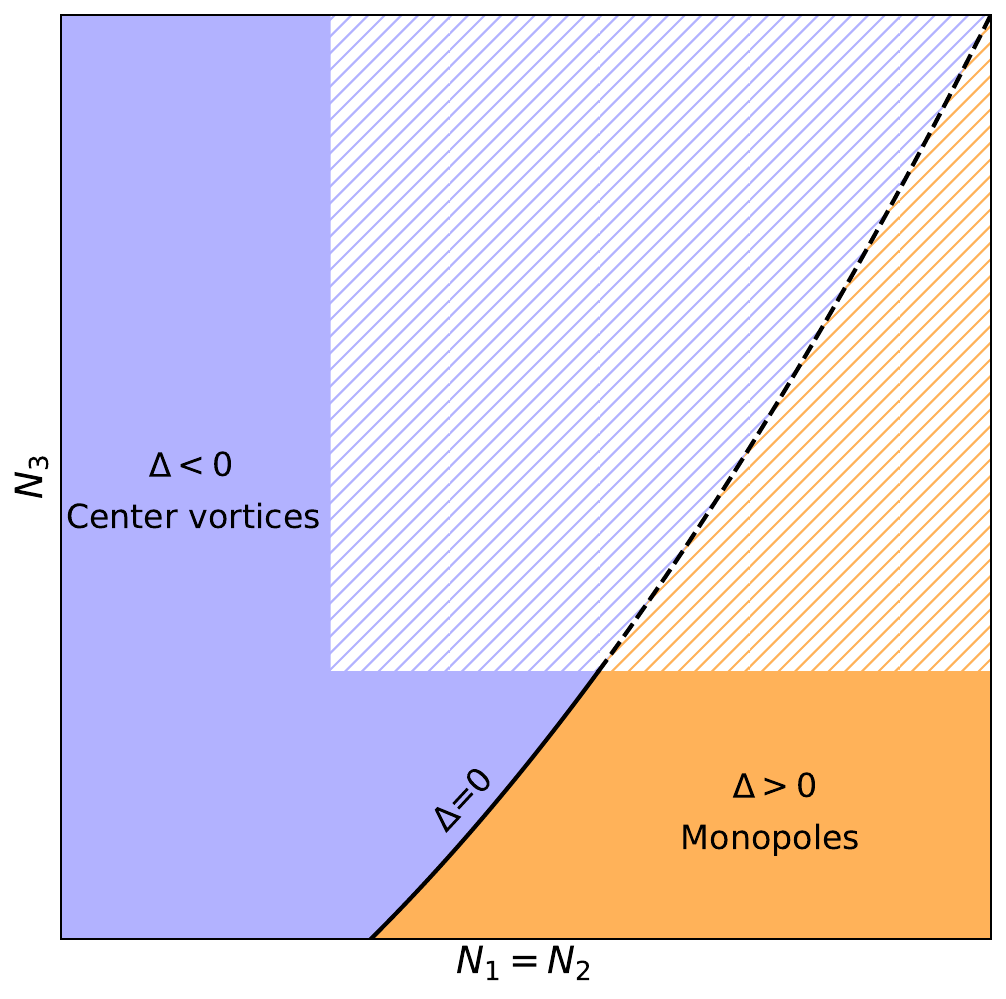}
    \caption{"Phase" diagram of the fractional instantons found through the cooling algorithm while varying $N_1$, $N_2$, and $N_3$ with $N_0$ kept large and fixed. The solid shaded regions and solid lines represent the regions of parameter space we explicitly test in this work. The hatched regions and dashed lines represent the expected continuation of these results throughout the rest of parameter space.}
    \label{fig:phase_diagram}
\end{figure}

The parameter space considered and our broad strokes findings are sketched in Figure \ref{fig:phase_diagram}. The main result is that the instantons are separated into two distinct classes: center vortex-like solutions when $\Delta<0$ and monopole-instanton-like solutions when $\Delta>0$. The key feature distinguishing these regimes is the asymptotic vacuum away from the localized excitations. Recall from Section \ref{sec:center_vortices} that the center vortex solutions are surrounded by vacua satisfying 
\begin{equation}
\label{eqn:12vacua}
    \begin{split}
    \frac{1}{2}\Tr W_1 = & \frac{1}{2}\Tr W_2 = \frac{1}{2}\Tr W_1W_2 = \,0 \\
    \frac{1}{2}\Tr W_3 = & \pm 1 \\
    \frac{1}{2}\Tr W_0 = & \pm 1 \, .
    \end{split}
\end{equation}
These asymptotic vacua are also found around all the configurations in the region $\Delta<0$. Recall also that these are the vacuum states we find when the $03$-twist is turned off.

In contrast, like the monopole-instantons considered in Section \ref{sec:monopoles}, all the fractional instantons in the region with $\Delta>0$ are surrounded by asymptotic vacua characterized by
\begin{equation}
\label{eqn:03vacua}
    \begin{split}
    \frac{1}{2}\Tr W_1 = & \pm 1 \\
    \frac{1}{2}\Tr W_2 = & \pm 1 \\
    \frac{1}{2}\Tr W_3 = & \frac{1}{2}\Tr W_0 = \frac{1}{2}\Tr W_0W_3 = \,0 \, .
    \end{split}
\end{equation}
By symmetry, these should be the vacuum states when considering the torus with only a $03$-twist (i.e. if we turn off the twists in the $12$-plane).

When both twists are turned on, it may asymptotically approach either the vacua in \ref{eqn:12vacua} or \ref{eqn:03vacua}, but because of the twists, the parameters $W_i$ must all be antiperiodic in one direction. Consider the case that that our configuration has the asymptotic vacua \ref{eqn:12vacua}. Then $\frac{1}{2}\Tr W_0$ must vary between $-1$ and $+1$ across the 3-direction and $\frac{1}{2}\Tr W_3$ must vary between $-1$ and $+1$ across the 0-direction. If the vacua are asymptotically \ref{eqn:03vacua}, then $\Tr W_1$ and $\Tr W_2$ will similarly vary across the $12$-plane. There is an energy cost associated with these variations, and there is a larger energy cost if the variations occur across a smaller plane. Therefore, if the $03$-plane is smaller than the $12$-plane (i.e. positive $\Delta$), it is energetically favourable for the asymptotic vacua to be \ref{eqn:03vacua}. Similarly, the asymptotic vacua in \ref{eqn:12vacua} are energetically favourable if the $12$-plane is smaller (i.e. negative $\Delta$). This heuristic argument agrees with the transition we see in the numerical data. 

The above argument raises the question of which vacuum wins when $\Delta=0$. This case is known as the symmetric torus, and the continuum theory was studied in depth by 't Hooft in \cite{tHooft:1981nnx}. He found analytical solutions for the BPS states. Importantly, these states have constant field strength and, therefore, constant action density. This implies that there is no asymptotic vacuum in this case, so neither vacuum wins. 

The overall picture is that the monopole-instantons spread out as $\Delta$ is lowered toward zero, becoming maximally spread out at $\Delta=0$, then the qualitatively different vortex-like solutions are formed as $\Delta$ becomes negative. This is reminiscent of a phase transition, though this is not sufficient evidence to claim a phase transition occurs at $\Delta=0$ for the full quantum system. 

It is important to note that this hard transition between vortex-like and monopole-like fractional instantons is a consequence of our particular path through parameter space and not a fundamental incompatibility between the two solutions. For example, in \cite{hayashi2024unifying}, they consider Yang-Mills on $\R^2\times S^1_{L_3} \times S^1_{L_4}$ with a twist in the $S^1_{L_3} \times S^1_{L_4}$. By keeping $L_4$ small and varying $L_3$, they argue that the monopole-instanton solutions at large $L_3$ smoothly deform to center vortices at small $L_3$. Their work focused on the magnetic dual description of the effective theory on $\R^2\times S^1_{L_3}$. Our algorithm is the ideal tool to provide evidence of this smooth transition in the full 4d electric description, however, such an investigation is beyond the scope of the current article.

In the following subsections, we present the configurations and discuss their properties in each of the regions $\Delta<0$, $\Delta>0$, and $\Delta\approx 0$. 

\subsection{$\Delta<0$}
In this region, the fractional instantons behave like the center vortices considered in Section \ref{sec:center_vortices}. 

\begin{figure}
    \centering
    \includegraphics[width=0.5\textwidth]{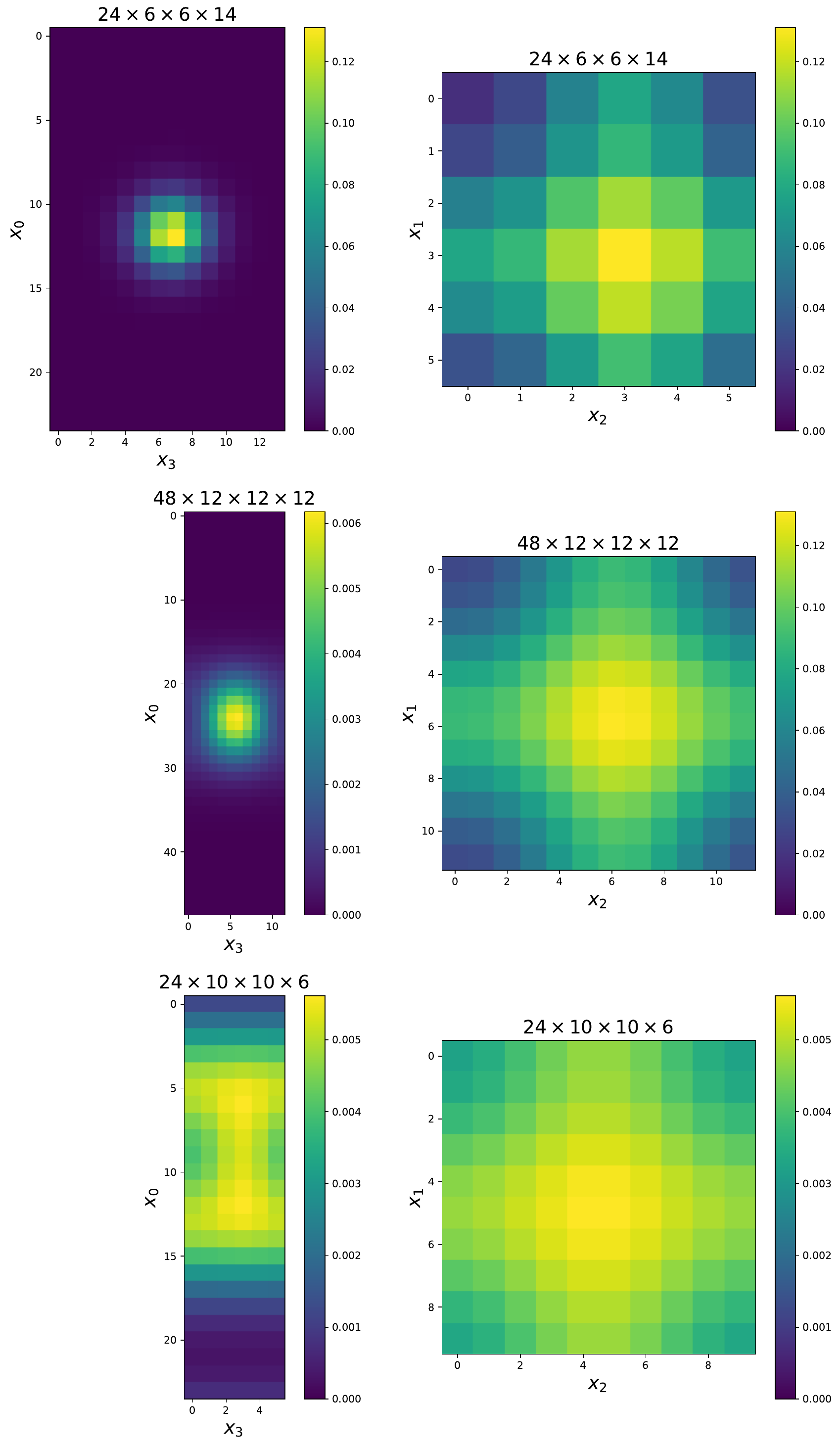}
    \caption{Plots of action densities in configurations of 3 different geometries with $\Delta<0$. The left hand column plots the $03$-plane that intersects the peak in action density and the right hand column plots the $12$-plane that intersects the peak in action density.}
    \label{fig:SDen_Dlt0}
\end{figure}

Consider the shape of the action density plots in Figure \ref{fig:SDen_Dlt0}. In agreement with the center vortices in the $\RT$ limit, there is localization in all 4 directions for all configurations in the $\Delta<0$ regime, though the localized excitation spreads out as $\Delta$ approaches 0. Interestingly, some additional structure appears when $\Delta\approx0$ (see the bottom right plot in Figure \ref{fig:SDen_Dlt0}). The nature of this additional structure is not entirely clear, but it remains when the configurations are generated from different initial seeds. Thus, it is a robust feature of these configurations. Exploring this structure further is beyond the scope of this article, but could be a rewarding exercise. This structure should also be accessible analytically through the $\Delta$-expansion presented in \cite{GarciaPerez:2000aiw}.

\begin{figure}
    \centering
    \includegraphics[width=0.7\textwidth]{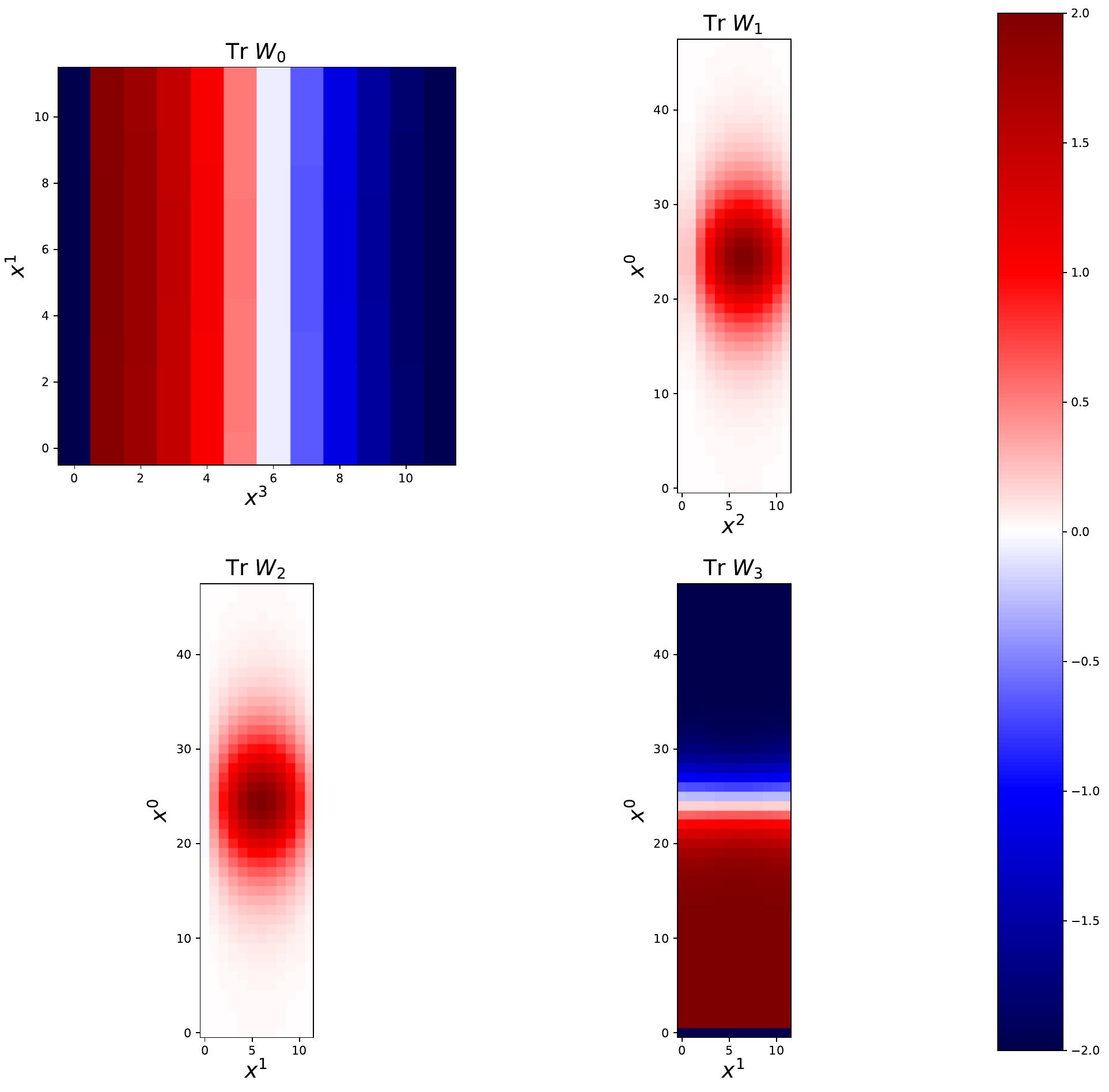}
    \caption{Traces of the holonomies of a lattice configuration with $\Delta<0$, namely $48\times12\times12\times12$. Holonomies are plotted along planes that intersect the point of maximal action density. Discontinuities are due to the fact that the holonomies are antiperiodic across twists, which are always contained in the plaquettes at the origin.}
    \label{fig:holonomy_Dlt0}
\end{figure}

To confirm that these configurations are center vortices, consider the holonomies of a representative example as shown in Figure \ref{fig:holonomy_Dlt0}. This Figure considers the "femtouniverse" configuration where all three spatial directions are kept small, but the results are characteristic of the qualitative structure of the holonomies throughout the $\Delta<0$ region of parameter space. The holonomies in Figure \ref{fig:holonomy_Dlt0} confirm that the asymptotic vacua agree with \ref{eqn:12vacua}. Moreover, they also confirm that these configurations work like center vortices in the $03$-plane since the $0$- and $3$-holonomies change sign when passing the configurations. 

Recall that the vacua will abelianize the theory to an effective 2D $\Z_2$ gauge theory. Since this theory has no propagating degrees of freedom and the gauge field is constrained to be $\Z_2$, the only real observable available is the $3$-holonomy and it only takes the values $\pm1$. Hence, perturbatively we would expect there to be two degenerate ground states $\ket{W_3=+1}$ and $\ket{W_3=-1}$. This degeneracy spontaneously breaks the center symmetry acting on loops that wind around the $3$-direction. It appears that our center vortex-like fractional instantons are exactly the instantons that interpolate between these classical vacua. Armed with these instantons and evidence that no other effects exist to cancel their effects, we could conclude that the instanton effects would split the degeneracy and leave a single center-symmetric ground state. This restores center symmetry and therefore confirms that the theory is in the confining phase.

However, the interpretation of our solutions as instantons interpolating between $\ket{W_3=+1}$ and $\ket{W_3=-1}$ is a little hasty. Recall that we originally found 4 classical vacua, based on the values of $W_3$ and $W_0$, and that our solutions interpolate between different signs of $W_0$ across the $3$-direction. When considering the $0$-direction as Wick rotated time, $W_0$ isn't a good observable on Hilbert space (which is defined on a given timeslice); however, it might be that another observable exists to distinguish these vacua. 

\begin{figure}
    \centering
    \includegraphics[width=0.7\textwidth]{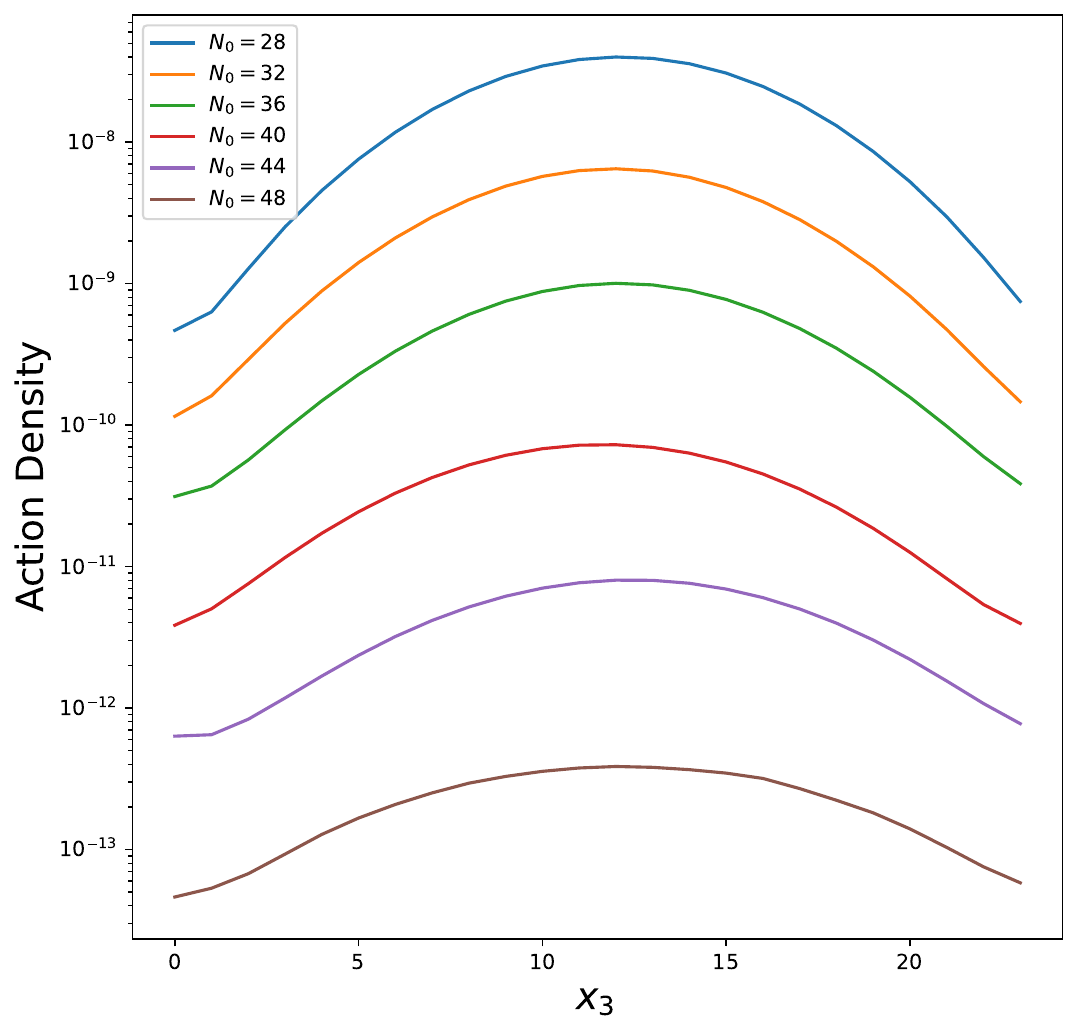}
    \caption{Plots showing that the apparent domain wall vanishes as $N_0$ grows. The action density is plotted at $x^0=1$ with $x^1$ and $x^2$ coinciding with the point of maximal action density. All configurations considered here are $N_0\times6\times6\times24$. Notice that as $N_0$ is increased the value at the peak of these distributions decreases by orders of magnitude and the distribution appears to flatten out.}
    \label{fig:DWcheck}
\end{figure}

If the asymptotic states at $x^0=1$ and $x^0=N_0$ interpolate between two vacua across the $3$-direction, then we would expect a domain wall to separate these vacua. Otherwise, if the asymptotic states are good approximations of true vacua, we should not see any domain wall forming. Such a domain wall would look like a localized bump in the action density along the $3$-direction at $x^0=1$. Plotting this, it appears that a domain wall exists (see the top line in Figure \ref{fig:DWcheck}), extinguishing our hopes of using center vortices in a simple instanton argument for confinement; however, we are concerned with the zero temperature limit, where we take $N_0$ to infinity before the other directions. Extending the $0$-direction and plotting the action density for different $N_0$, we find that the peak of the "domain wall" quickly decays away as demonstrated in Figure \ref{fig:DWcheck}. This suggests that in the zero temperature limit, the center vortex solutions in this regime indeed interpolate between $\ket{W_3=+1}$ and $\ket{W_3=-1}$, restoring center-symmetry and confinement. This agrees with the analysis of instantons in \cite{GARCIAPEREZ:1993}. 

Thus, there is a single cohesive confinement mechanism (based on center-vortices) for YM on every $\RTS$ with a sufficiently small $\T^2$ to force abelianization to $\Z_2$. Given the basis on center vortices, this mechanism is also consistent with the confining mechanism in the semi-infinite volume limit, $\RT$.

\subsection{$\Delta>0$}
\begin{figure}
    \centering
    \includegraphics[width=0.5\textwidth]{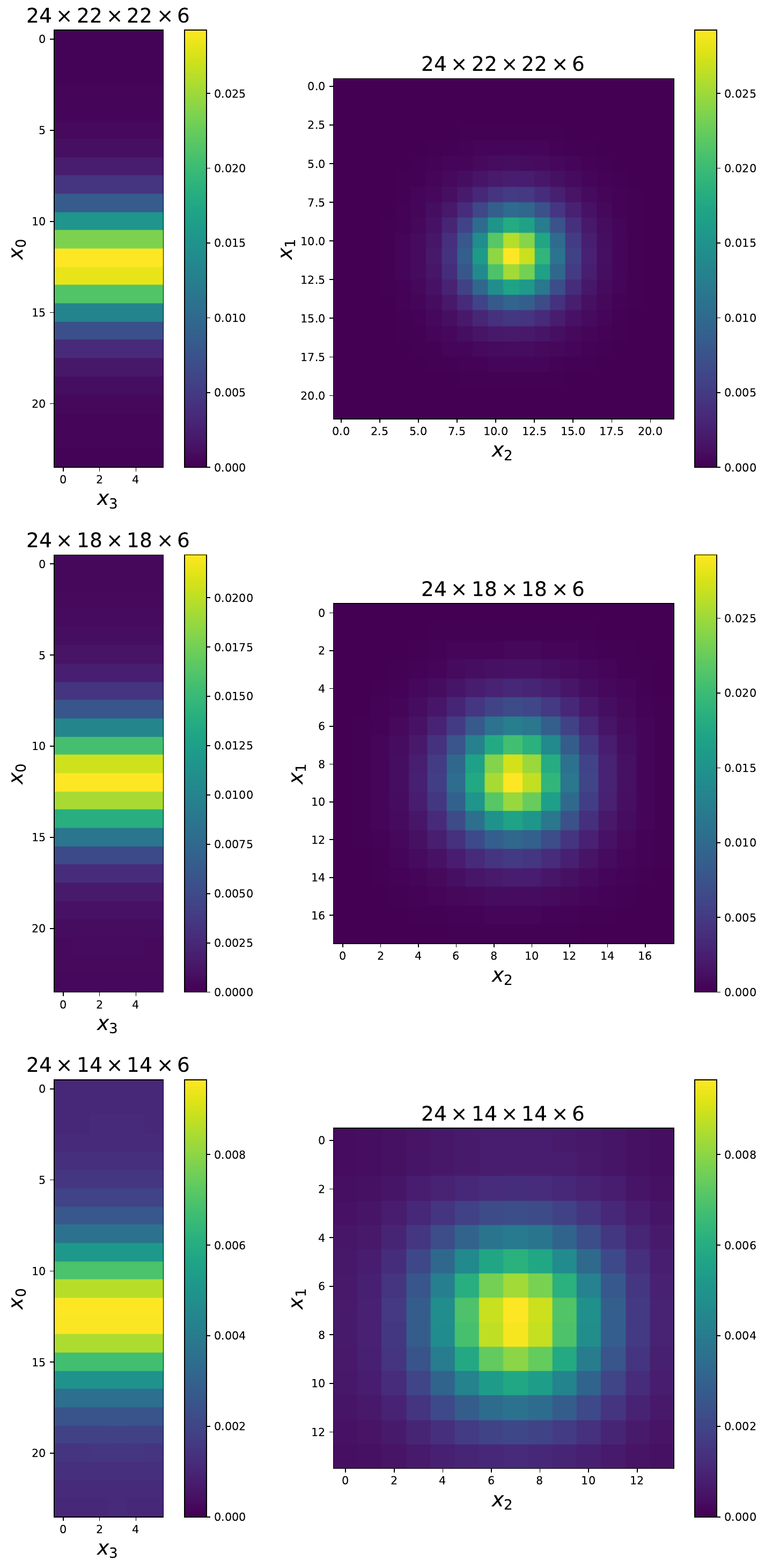}
    \caption{Plots of action densities in configurations of 3 different geometries with $\Delta>0$. The left hand column plots the $03$-plane that intersects the peak in action density and the right hand column plots the $12-plane$ that intersects the peak in action density.}
    \label{fig:SDen_Dgt0}
\end{figure}

In this region, the instantons are monopoles like those described in Section \ref{sec:monopoles}. Figure \ref{fig:SDen_Dgt0} shows the action density of configurations in this regime. Notice that the configurations are completely spread out along the $3$-direction with no peak, just like the monopole solutions shown in Figure \ref{fig:SDen_monopole}. The monopoles spread out as $\Delta$ gets closer to 0 (top to bottom in Figure \ref{fig:SDen_Dgt0}). 

\begin{figure}
    \centering
    \includegraphics[width=0.7\textwidth]{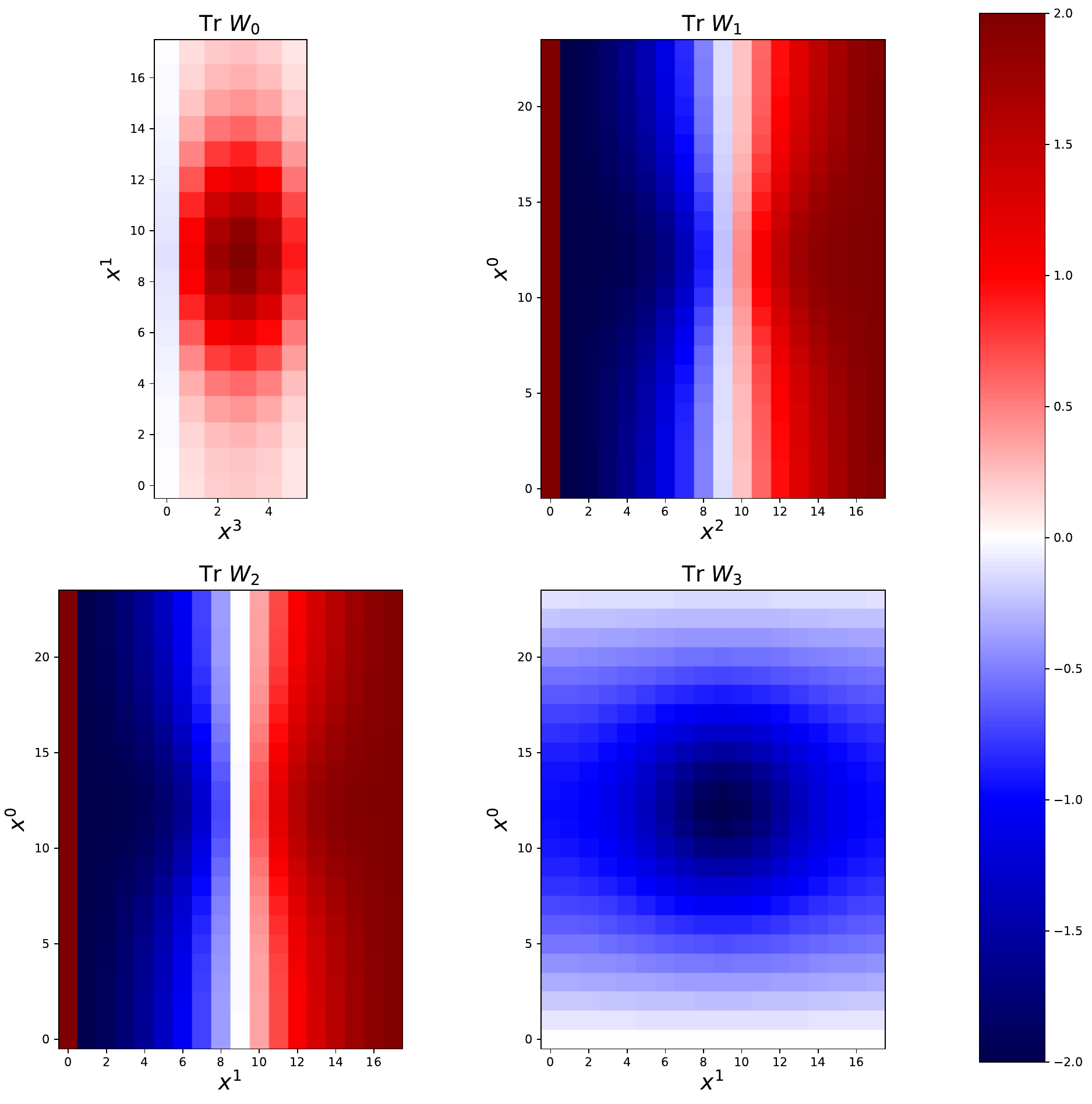}
    \caption{Traces of the holonomies of a lattice configuration with $\Delta>0$, namely $24\times18\times18\times6$. Holonomies are plotted along planes that intersect the point of maximal action density. Discontinuities are due to the fact that the holonomies are antiperiodic across twists, which are always contained in the plaquettes at the origin.}
    \label{fig:holonomy_Dgt0}
\end{figure}

Figure \ref{fig:holonomy_Dgt0} shows the holonomies for a representative configuration with $\Delta>0$. Here we can see that the asymptotic vacua agree with \ref{eqn:03vacua}. As mentioned in Section \ref{sec:monopoles}, this raises difficulties with using these configurations for semiclassics in pure YM, but it also allows for us to calculate the magnetic fields surrounding these instantons using Equation \ref{eqn:Bdef}.

\begin{figure}
    \centering
    \includegraphics[width=0.9\textwidth]{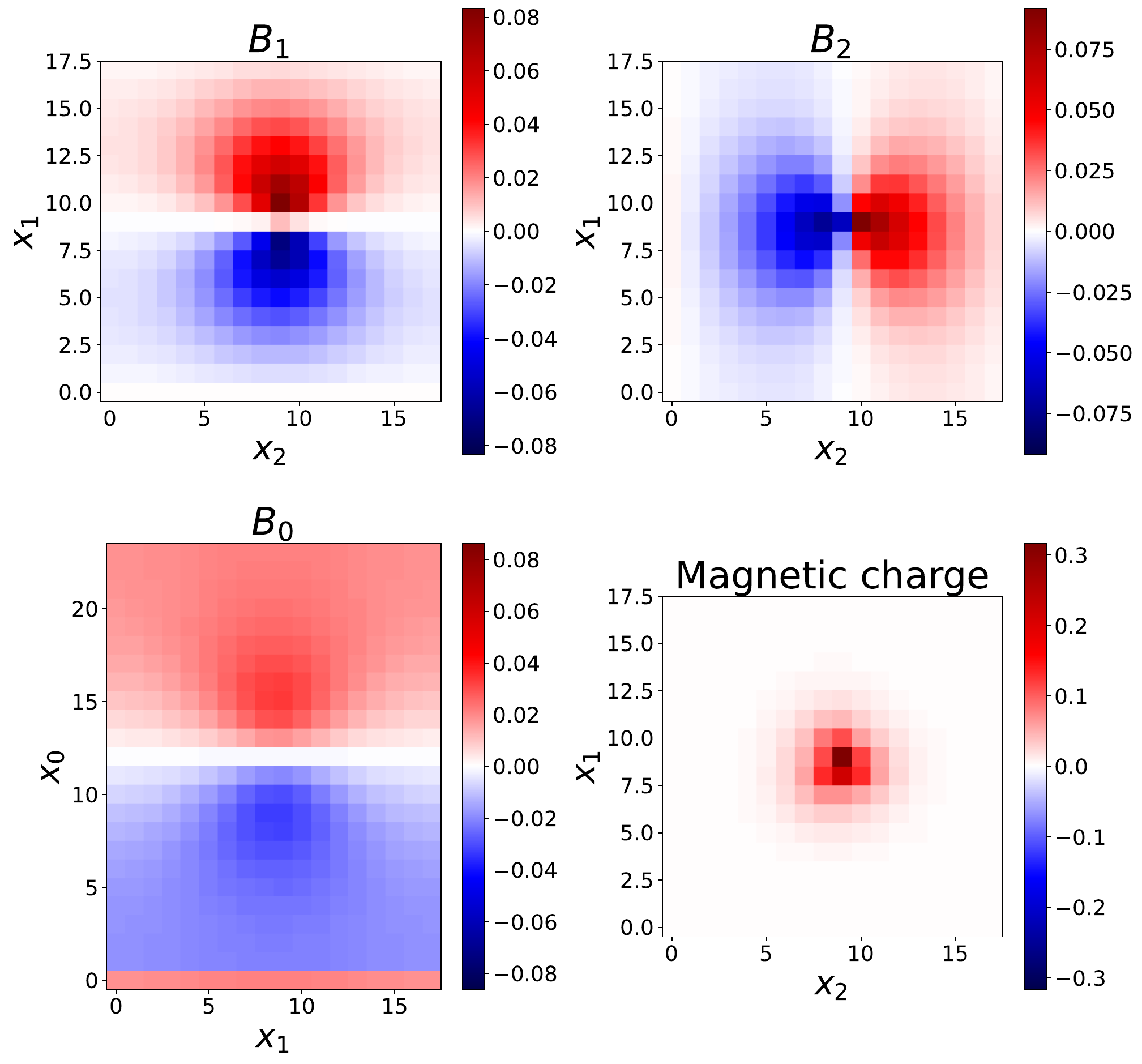}
    \caption{Magnetic field for the lattice configuration with $24\times18\times18\times6$. Notice the qualitative similarity to Figure \ref{fig:Bfield}.}
    \label{fig:Bfield_Dgt0}
\end{figure}

The magnetic fields of a characteristic configuration within the $\Delta>0$ region are plotted in Figure \ref{fig:Bfield_Dgt0}. There are clear similarities with the magnetic fields in Figure \ref{fig:Bfield}. To demonstrate that these are truly monopoles, we test the quantization of magnetic charge by measuring the total flux. The results are shown in Table \ref{tab:fluxes_Dgt0}. All the values are consistent with the fluxes measured for similarly sized lattices in Section \ref{sec:monopoles}. As demonstrated previously, these values are consistent with having a $4\pi$ flux in the continuum limit. This is the flux required for continuum monopoles. 

\begin{table}[]
    \centering
    \begin{tabular}{| c | c | c|}
         \hline
         $N_1=N_2$ & $\Delta$ & Total magnetic flux ($\Phi_3$) \\
         \hline
         14 & 0.3095 & 12.5718\\
         16 & 0.5833 & 12.5680\\
         18 & 0.8333 & 12.5665\\
         20 & 1.0666 & 12.5665\\
         22 & 1.2878 & 12.5670 \\
         24 & 1.5 & 12.5662\\
         \hline
    \end{tabular}
    \caption{Values of the total magnetic flux, $\Phi_3$, in the region $\Delta>0$. $\Delta$ is varied by changing $N_1$ and $N_2$ while maintaining $N_1=N_2$. The values of $N_0$ and $N_3$ are fixed at 24 and 6, respectively.}
    \label{tab:fluxes_Dgt0}
\end{table}


While the structure of the monopoles is interesting, it remains difficult to use them for semiclassics in YM on $\RTS$. First, there is the issue of the asymptotic vacuum differing from the true vacuum. Moreover, the true vacuum does not abelianize the theory. These were already mentioned in Section \ref{sec:monopoles} where we discussed the $\RS$ limit. In the $\RTS$ limit, there is another issue: $\Delta$ is never positive in the zero temperature limit. Recall that the zero temperature limit involves taking $N_0\rightarrow\infty$ first. As $N_0$ grows, $\Delta$ decreases and eventually turns negative. Hence, there is no 4-dimensional zero temperature limit\footnote{One could consider keeping $\T^2_*$ finite and $\Delta$ positive by making the $S^1$ vanishingly small, but the continuum  limit in this case would be the dimensional reduction on $\R\times\T^2_*$. This might be interesting for studying the twisted Polyakov model, but it is not useful for the present article's goal of studying the 4-dimensional case.} consistent with a finite $\T^2_*$ and $\Delta>0$.

It is the author's hope that all these difficulties can be overcome with the introduction of a deformation potential that stabilizes the center symmetry in the 3-direction. This deformation potential should make a vacuum with $\Tr W_3=0$ energetically preferred and should hopefully stabilize monopole instantons with similar qualitative features within the $\Delta<0$ region. This should lead to another cohesive picture of confinement on $\RTS$ whenever the $\T^2_*$ is too large to abelianize down to $\Z_2$, but the $\S^1$ is small enough to abelianize to $U(1)$. However, as mentioned before, numerics in the presence of the deformation potential are significantly more involved, so the inclusion of this potential is outside the scope of the present article.

\subsection{$\Delta=0$}
We now understand that the sign of $\Delta$ determines whether the BPS fractional instantons in the theory are monopoles or center vortices, but this leaves the question of what happens at $\Delta=0$. In our simulations, this corresponds to a lattice   size of $24\times12\times12\times6$. This is analogous to the case of the self-dual torus, where we have analytical continuum solutions due to 't Hooft \cite{tHooft:1981nnx}.

To see the solutions consider a continuum $\T^4$ with lengths $L_0$, $L_1$, $L_2$, and $L_3$ in each direction. Here $\Delta$ is defined as
\begin{equation}
    \Delta = \frac{L_1L_2-L_0L_3}{\sqrt{L_0L_1L_2L_3}}\, .
\end{equation}
Hence, $\Delta=0$ implies $L_1L_2 = L_0L_3$. In the continuum, the $03$-plane and $12$-plane twists are most easily implemented with boundary  conditions. We denote the boundary conditions by $\Omega_\mu(x)$ such that 
\begin{equation}
    A(x + L_\mu e_\mu) = \Omega_\mu (A-id) \Omega_\mu^\dag\, .
\end{equation}
The following boundary conditions implement the desired twist:
\begin{equation}
    \begin{split}
        \Omega_0(x) = & e^{-i\pi \frac{x^3}{L_3}\sigma^3}\\
        \Omega_1(x) = & 1\\
        \Omega_2(x) = & e^{-i\pi \frac{x^1}{L_1}\sigma^3}\\
        \Omega_3(x) = & 1 \, .
    \end{split}
\end{equation}
With these boundary conditions, 't Hooft's self-dual solutions at $\Delta=0$ are easy to write down:
\begin{equation}
\label{eqn:thooftsolution}
    \begin{split}
        A_0 = & \frac{z_0}{L_0} \frac{\sigma^3}{2} \\
        A_1 = & \left(\frac{2\pi x^2}{L_1L_2}  + \frac{z_1}{L_1}\right)\frac{\sigma^3}{2}\\
        A_2 = & \frac{z_2}{L_2} \frac{\sigma^3}{2} \\
        A_3 = & \left(\frac{2\pi x^0}{L_0L_3}  + \frac{z_3}{L_3}\right)\frac{\sigma^3}{2} \,  .
    \end{split}
\end{equation}
Here the $z_i$ are dimensionless parameters that parameterize the moduli space of these solutions. 

In this gauge, the field strength tensor is constant, taking the values
\begin{equation}
\begin{split}
    F_{03} = -F_{30} = & \frac{2\pi}{L_0L_3} \frac{\sigma^3}{2}\\
    F_{12} = -F_{21} = & \frac{2\pi}{L_1L_2} \frac{\sigma^3}{2}\\
\end{split}
\end{equation}
with all other components identically zero. The constant field strength is not a gauge invariant property, but it guarantees that gauge invariant observables built out of the field strength, such as the action density, are  constant. In  other words, the action density plots should show that the $\Delta=0$ solitons are maximally spread out. 
 
\begin{figure}
    \centering
    \includegraphics[width=0.5\textwidth]{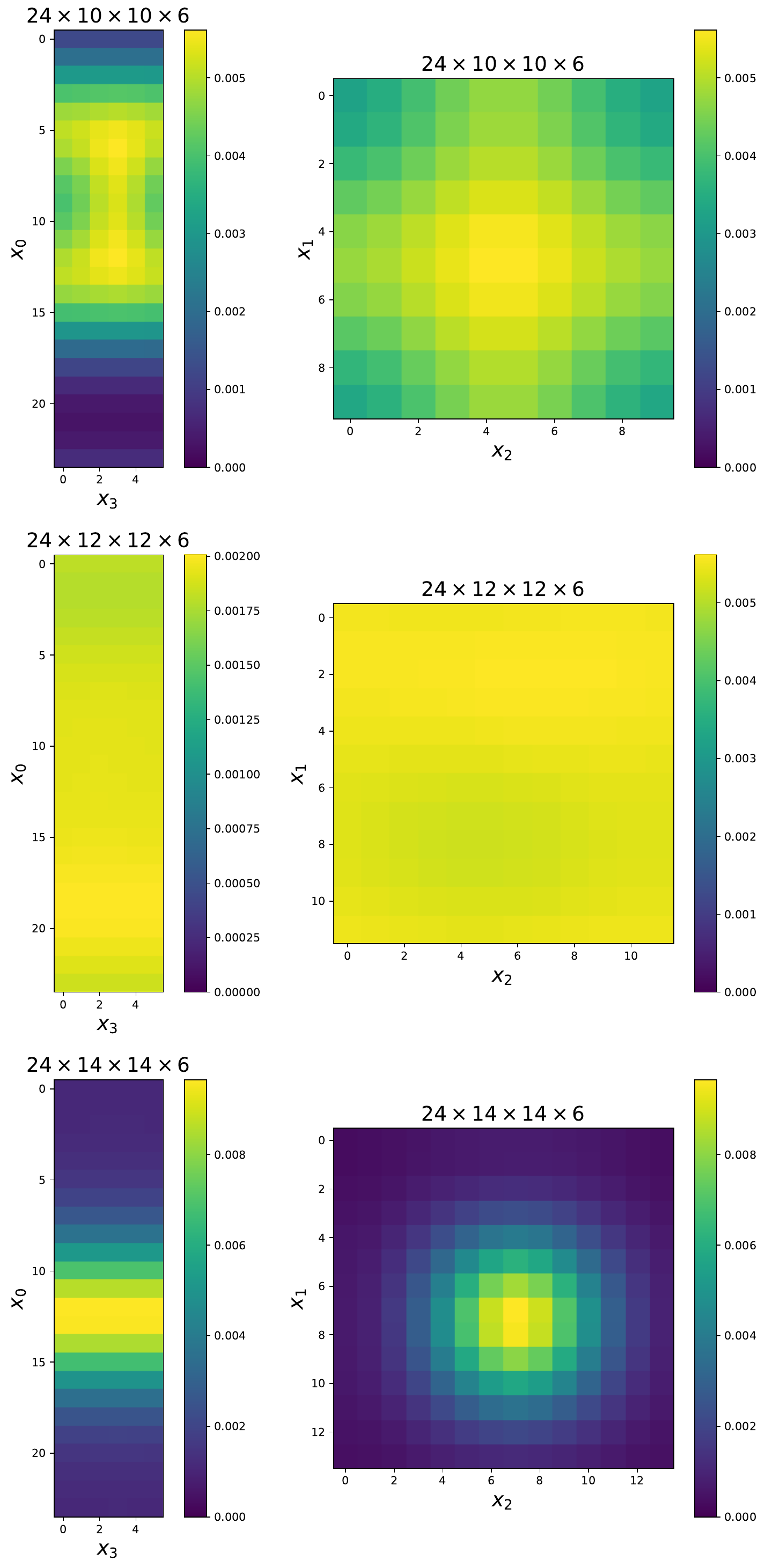}
    \caption{Plots of the action density in configurations with $\Delta=0$ and near $\Delta=0$. The $24\times12\times12\times6$ configuration has exactly $\Delta=0$. The other configurations are included for comparison. As with all previous action density plots, the left hand column plots the $03$-plane that intersects the peak in action density and the right hand column plots the $12-plane$ that intersects the peak in action density.}
    \label{fig:SDen_nearD0}
\end{figure}

Figure \ref{fig:SDen_nearD0} displays such plots. It is clear from the figure that the action density is not completely uniform, there is a peak in each plane. However, this peak is very small compared to the baseline value of the action density.

To gain a quantitative understanding of how much the action deviates from a constant solution, we compare the average value of the action density with the maximum, minimum and standard deviations. The results are collected in Table \ref{tab:average_SDens}, along with a lattice configuration of size $24\times14\times14\times6$ and $24\times10\times10\times6$. The data clearly demonstrates that the solution is very close to uniform, especially when compared to the monopole and center vortex solutions. This suggests that we are approximating the continuum constant field strength solutions reasonably well, so we can reasonably blame the deviation away from the constant solution to the effects of finite lattice spacing.

\begin{table}
    \centering
    \begin{tabular}{|c|cc||cc||cc|}
    \hline
         & \multicolumn{2}{c||}{$24\times12\times12\times6$} & \multicolumn{2}{c||}{$24\times14\times14\times6$} & \multicolumn{2}{c|}{$24\times10\times10\times6$} \\
         \cline{2-7}
        & Value & \% error & Value & \% error & Value & \% error\\
        \hline
        Mean & 0.001904 & & 0.001395 & & 0.002734 & \\
        Standard deviation & 0.000070 & 3.7\% & 0.001154 & 82.7\% & 0.001573 & 57.5\% \\
        Minimum & 0.001691 & -11.2\% & 0.000333 & -76.2\% & 0.000301 & -89.0\% \\
        Maximum & 0.002009 & 5.5\% & 0.009651 & 591.8\% & 0.005614 & 105.4\% \\
        \hline
    \end{tabular}
    \caption{Average and variation of the action density across configurations with different geometries. The percent errors are calculated with respect to the mean action density within the same geometry. The key point is that the action density on the self-dual torus (i.e. the $24\times12\times12\times6$ lattice) varies little compared to the mean, especially compared to fractional instantons on lattices with $\Delta\neq0$.}
    \label{tab:average_SDens}
\end{table}

It is also interesting to consider what happens as $\Delta$ approaches 0 from both sides. Figure \ref{fig:SDen_nearD0} demonstrates how both the center vortices and the monopoles spread out at small values of $\left|\Delta\right|$. Here it is clear that the transition between the two regimes is sharp, since the unique shapes of the configurations are recognizable in all cases except exactly at $\Delta=0$. 

Considering Equation \ref{eqn:thooftsolution} and including the boundary conditions, we can also work out the holonomies in this solution:
\begin{equation}
\label{eqn:holonomy_Deq0}
    \begin{split}
        W_0 = \, & e^{i\left(z_0 - 2\pi\frac{x^3}{L_3}\right)\frac{\sigma^3}{2}}\\
        W_1 = \, & e^{i\left(z_1 + 2\pi\frac{x^2}{L_2}\right)\frac{\sigma^3}{2}}\\
        W_2 = \, & e^{i\left(z_2 - 2\pi\frac{x^1}{L_1}\right)\frac{\sigma^3}{2}}\\
        W_3 = \, & e^{i\left(z_3 + 2\pi\frac{x^0}{L_0}\right)\frac{\sigma^3}{2}} \, .
    \end{split}
\end{equation}
It is clear from these equations that shifting the $z_i$ can be reabsorbed into a shift of the coordinates $x^i$. Hence, the moduli space consists of translations, just as for the other configurations we studied. It also means we can shift the solutions such that the $z_i$ vanish.

\begin{figure}
    \centering
    \includegraphics[width=0.7\textwidth]{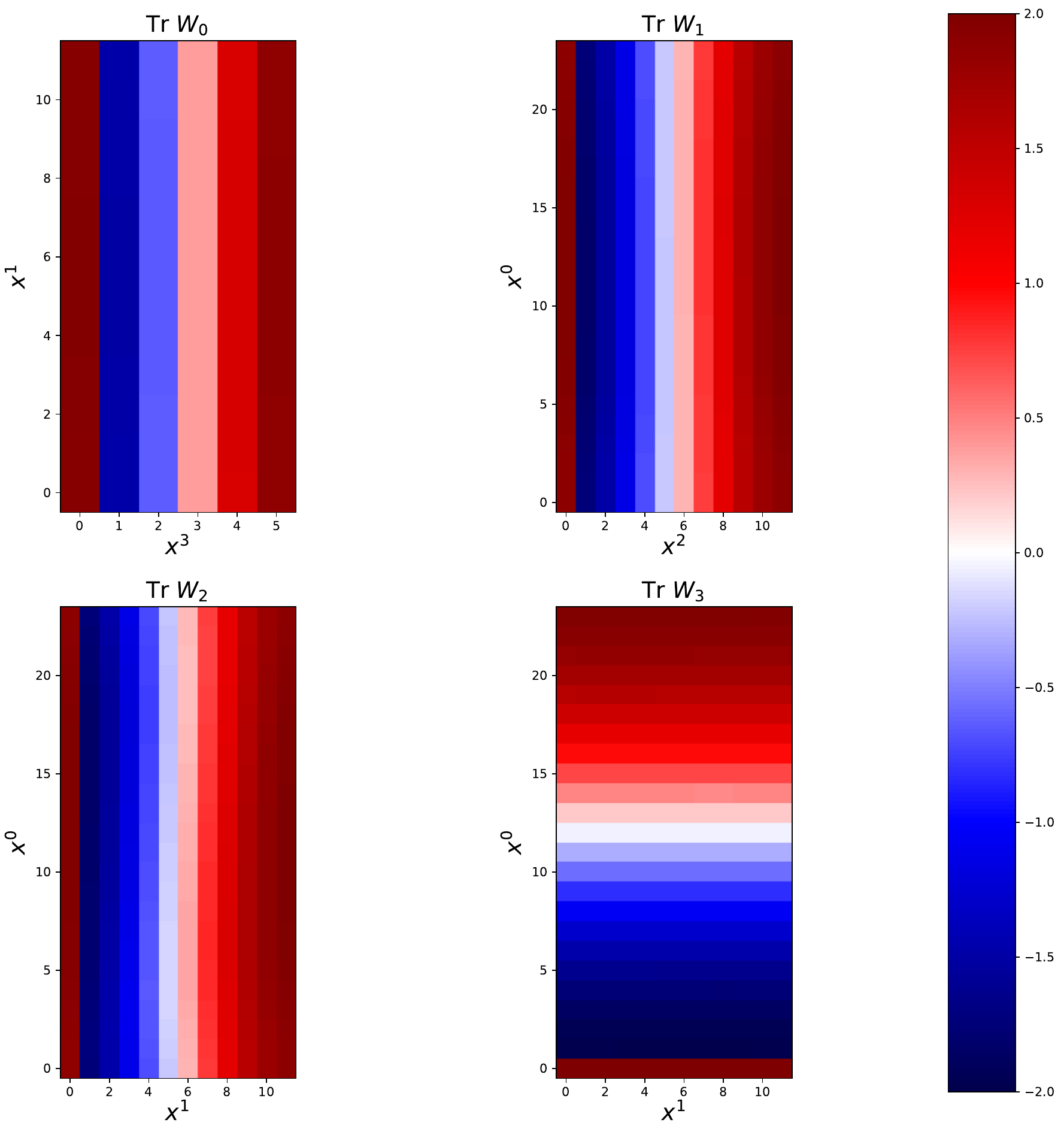}
    \caption{Traces of the holonomies of the lattice configuration with $\Delta=0$, namely $24\times12\times12\times6$. By comparison with Equation \ref{eqn:holonomy_Deq0}, it is clear that this configuration has been translated such that the $z_\mu$ vanish. Holonomies are plotted along planes that intersect the point of maximal action density. Discontinuities are due to the fact that the holonomies are antiperiodic across twists, which are always contained in the plaquettes at the origin.}
    \label{fig:holonomy_Deq0}
\end{figure}

Considering Equation \ref{eqn:holonomy_Deq0} with $z_i=0$. In this case, the gauge invariant traces of the holonomies in each of the four directions should interpolate between $-1$ and $+1$ across some other direction. We can see the agreement with this in Figure \ref{fig:holonomy_Deq0}. This gives further evidence that we are approximating 't Hooft's solutions with our numerical solutions, and suggests that the moduli space on the self-dual torus agrees with Equation \ref{eqn:modspace}.

\section{Conclusion and future directions}
\label{sec:conclusions}
We used a numerical cooling algorithm to investigate the fractional instantons that exist on $\RT$, $\RS$, and $\RTS$. We considered lattices with sizes $N_0 \times N_1\times N_2\times N_3$. For all configurations, $N_0$ was kept large to approximate Euclidean time at low temperature, $N_1$ and $N_2$ were kept equal and a twist included in the $12$-plane to make an approximate $\T^2_*$, and $N_3$ was varied to approximate different sizes of the $S^1$ direction. $N_1$, $N_2$, and $N_3$ were varied at fixed $N_0$ to approximate the various continuum limits. To find fractional instantons we included an additional twist in the $03$-plane to force a fractional topological charge. The parameter
\begin{equation}
    \Delta = \frac{N_1N_2 - N_0N_3}{\sqrt{N_0N_1N_2N_3}}
\end{equation}
was useful for distinguishing different regimes in the parameter space.

The solutions came in three varieties: center vortices when $\Delta$ was negative, monopole instantons when $\Delta$ was positive, and 't Hooft's uniform solutions when $\Delta$ exactly vanished. We confirmed that these identifications were accurate by calculating characteristic gauge invariant quantities in each regime, such as confirming linking with Wilson loops for the center vortices and correct magnetic flux quantization in the monopole instantons.

This numerical evidence demonstrates that the center vortices have the correct properties to ensure confinement on $\RT$ and $\RTS$ whenever the $\T^2_*$ is sufficiently small. This strengthens previous numerical evidence and expands the validity to a larger class of geometries, namely $\RTS$ with any size of circle. 

While the monopole instantons require additional center stabilization to be effective in explaining confinement, we showed that there is a sharp transition between monopoles instantons on $\RS$ and the center vortices on $\RTS$. If we allow for finite temperature, then we find monopole instantons on $S^1_\beta\times\T^2_*\times S^1_L$ for large $\T^2_*$ and the sharp transition occurs at some finite size of $\T^2_*$. The fractional instantons at the point of transition are given by 't Hooft's constant field strength solutions. The details of this sharp transition are, as far as the author knows, new to this work.

There is rich physics hiding in these classical solutions and their semiclassical effects on the full quantum theory. Several extensions and further uses for the algorithm were discussed in the text. One promising direction that would not require any modifications to the algorithm would be to investigate the electric description of the setup from \cite{hayashi2024unifying}, which should have a smooth crossover between the monopole instantons and the center vortices, in contrast to the sharp transition we found. This investigation could be carried out by changing the twists to live in the $01$- and $23$-planes, and investigating configurations of the form $N_L\times N_L\times N_I\times N_S$, where $N_L$ is a large value, $N_S$ is a small value, and $N_I$ interpolates between the large and small values. 

In addition, there might be some interesting physics to be found in the additional structure observed when $-1 \ll \Delta < 0$. The nature of this structure should agree with the analytical $\Delta$-expansion solutions on nearly self-dual torii developed in \cite{GarciaPerez:2000aiw}.

An important generalization of this algorithm would be the inclusion of a deformation potential. The deformation potential necessitates a more involved algorithm, but it would allow for a direct numerical test of the continuity between regimes discussed in \cite{Poppitz2023} and \cite{guvendik2024metamorphosis}. This is because the theory with deformation potential would energetically favour the $F_{12}\neq 0$ vacua these papers require to discuss semiclassics. Moreover, it should allow monopole instantons at zero temperature on $\RTS$ with finite $\T^2_*$. It might even smooth out the sharp transition we observed in pure YM theory, confirming the crossover hypothesis in \cite{guvendik2024metamorphosis}.

\bigskip

{\flushleft{\bf Acknowledgements:}} I would like to thank Erich Poppitz for hours of helpful discussion and for asking all the difficult questions that lead to the most interesting results in this work.

\bigskip

 \bibliography{SU2Sol.bib}

\providecommand{\href}[2]{#2}\begingroup\raggedright\begin{thebibliography}{10}

\bibitem{Poppitz:2021cxe}
E.~Poppitz, {\it {Notes on Confinement on $\R^3 \times \S^1$: From Yang\textendash{}Mills, Super-Yang\textendash{}Mills, and QCD (adj) to QCD(F)}},  {\em Symmetry} {\bf 14} (2022), no.~1 180, [\href{http://arxiv.org/abs/2111.10423}{{\tt arXiv:2111.10423}}].

\bibitem{Unsal2008adj}
M.~\"Unsal, {\it Abelian duality, confinement, and chiral-symmetry breaking in a su(2) qcd-like theory},  {\em Phys. Rev. Lett.} {\bf 100} (Jan, 2008) 032005.

\bibitem{Unsal2009bions}
M.~\"Unsal, {\it Magnetic bion condensation: A new mechanism of confinement and mass gap in four dimensions},  {\em Phys. Rev. D} {\bf 80} (Sep, 2009) 065001.

\bibitem{Shifman:2008ja}
M.~Shifman and M.~Unsal, {\it {QCD-like Theories on R(3) x S(1): A Smooth Journey from Small to Large r(S(1)) with Double-Trace Deformations}},  {\em Phys. Rev. D} {\bf 78} (2008) 065004, [\href{http://arxiv.org/abs/0802.1232}{{\tt arXiv:0802.1232}}].

\bibitem{Unsal:2008ch}
M.~Unsal and L.~G. Yaffe, {\it {Center-stabilized Yang-Mills theory: Confinement and large N volume independence}},  {\em Phys. Rev. D} {\bf 78} (2008) 065035, [\href{http://arxiv.org/abs/0803.0344}{{\tt arXiv:0803.0344}}].

\bibitem{Tanizaki:2022ngt}
Y.~Tanizaki and M.~\"Unsal, {\it {Center vortex and confinement in Yang\textendash{}Mills theory and QCD with anomaly-preserving compactifications}},  {\em PTEP} {\bf 2022} (2022), no.~4 04A108, [\href{http://arxiv.org/abs/2201.06166}{{\tt arXiv:2201.06166}}].

\bibitem{Oxman:2018prd}
L.~E. Oxman, {\it 4d ensembles of percolating center vortices and monopole defects: The emergence of flux tubes with $n$-ality and gluon confinement},  {\em Phys. Rev. D} {\bf 98} (Aug, 2018) 036018.

\bibitem{Oxman2018epjc}
L.~E. Oxman and H.~Reinhardt, {\it Effective theory of the d = 3 center vortex ensemble},  {\em The European Physical Journal C} {\bf 78} (Mar, 2018) 177.

\bibitem{Junior2020}
D.~R. Junior, L.~E. Oxman, and G.~M. Sim{\~o}es, {\it 3d yang-mills confining properties from a non-abelian ensemble perspective},  {\em Journal of High Energy Physics} {\bf 2020} (Jan, 2020) 180.

\bibitem{Junior2021}
D.~R. Junior, L.~E. Oxman, and G.~M. Simões, {\it From center-vortex ensembles to the confining flux tube},  {\em Universe} {\bf 7} (2021), no.~8.

\bibitem{Junior2022}
D.~R. Junior, L.~E. Oxman, and H.~Reinhardt, {\it Infrared yang-mills wave functional due to percolating center vortices},  {\em Phys. Rev. D} {\bf 106} (Dec, 2022) 114021.

\bibitem{Poppitz2023}
E.~Poppitz and F.~D. Wandler, {\it {Gauge theory geography: charting a path between semiclassical islands}},  {\em JHEP} {\bf 02} (2023) 014, [\href{http://arxiv.org/abs/2211.10347}{{\tt arXiv:2211.10347}}].

\bibitem{Perez:2013aa}
M.~Garc\'\i{}a~P\'erez, A.~Gonz\'alez-Arroyo, and M.~Okawa, {\it {Spatial volume dependence for 2+1 dimensional SU(N) Yang-Mills theory}},  {\em JHEP} {\bf 09} (2013) 003, [\href{http://arxiv.org/abs/1307.5254}{{\tt arXiv:1307.5254}}].

\bibitem{Unsal:2020yeh}
M.~\"Unsal, {\it {Strongly coupled QFT dynamics via TQFT coupling}},  {\em JHEP} {\bf 11} (2021) 134, [\href{http://arxiv.org/abs/2007.03880}{{\tt arXiv:2007.03880}}].

\bibitem{Wilson:1974sk}
K.~G. Wilson, {\it {Confinement of Quarks}},  {\em Phys. Rev. D} {\bf 10} (1974) 2445--2459.

\bibitem{Laursen1988}
M.~L. Laursen and G.~Schierholz, {\it Evidence for monopoles in the quantizedsu(2) lattice vacuum: A study at finite temperature},  {\em Zeitschrift f{\"u}r Physik C Particles and Fields} {\bf 38} (Sep, 1988) 501--509.

\bibitem{GarciaPerez:1989gt}
M.~Garcia~Perez, A.~Gonzalez-Arroyo, and B.~Soderberg, {\it {Minimum Action Solutions for SU(2) Gauge Theory on the Torus With Nonorthogonal Twist}},  {\em Phys. Lett. B} {\bf 235} (1990) 117--123.

\bibitem{GarciaPerez:1992fj}
M.~Garcia~Perez and A.~Gonzalez-Arroyo, {\it {Numerical study of Yang-Mills classical solutions on the twisted torus}},  {\em J. Phys. A} {\bf 26} (1993) 2667--2678, [\href{http://arxiv.org/abs/hep-lat/9206016}{{\tt hep-lat/9206016}}].

\bibitem{GONZALEZARROYO1998273}
A.~González-Arroyo and A.~Montero, {\it Self-dual vortex-like configurations in su(2) yang-mills theory},  {\em Physics Letters B} {\bf 442} (1998), no.~1 273--278.

\bibitem{Berg:1981nw}
B.~Berg, {\it {Dislocations and Topological Background in the Lattice O(3) $\sigma$ Model}},  {\em Phys. Lett. B} {\bf 104} (1981) 475--480.

\bibitem{Teper:1985rb}
M.~Teper, {\it {Instantons in the Quantized SU(2) Vacuum: A Lattice Monte Carlo Investigation}},  {\em Phys. Lett. B} {\bf 162} (1985) 357--362.

\bibitem{GPY1981}
D.~J. Gross, R.~D. Pisarski, and L.~G. Yaffe, {\it Qcd and instantons at finite temperature},  {\em Rev. Mod. Phys.} {\bf 53} (Jan, 1981) 43--80.

\bibitem{tHooft:1981nnx}
G.~'t~Hooft, {\it {Some Twisted Selfdual Solutions for the Yang-Mills Equations on a Hypertorus}},  {\em Commun. Math. Phys.} {\bf 81} (1981) 267--275.

\bibitem{hayashi2024unifying}
Y.~Hayashi and Y.~Tanizaki, {\it Unifying monopole and center vortex as the semiclassical confinement mechanism},  2024.

\bibitem{guvendik2024metamorphosis}
C.~Güvendik, T.~Schaefer, and M.~Ünsal, {\it {The metamorphosis of semi-classical mechanisms of confinement: From monopoles on $\R^3 \times \S^1$ to center-vortices on $\R^2 \times \T^2$}},  2024.

\bibitem{Bogomolny:1975de}
E.~B. Bogomolny, {\it {Stability of Classical Solutions}},  {\em Sov. J. Nucl. Phys.} {\bf 24} (1976) 449.

\bibitem{Prasad:1975kr}
M.~K. Prasad and C.~M. Sommerfield, {\it {An Exact Classical Solution for the 't Hooft Monopole and the Julia-Zee Dyon}},  {\em Phys. Rev. Lett.} {\bf 35} (1975) 760--762.

\bibitem{Gonzalez-Arroyo:1997ugn}
A.~Gonzalez-Arroyo, {\it {Yang-Mills fields on the four-dimensional torus. Part 1.: Classical theory}},  in {\em {Advanced Summer School on Nonperturbative Quantum Field Physics}}, pp.~57--91, 6, 1997.
\newblock \href{http://arxiv.org/abs/hep-th/9807108}{{\tt hep-th/9807108}}.

\bibitem{ruleof3}
C.~L. Rumke, {\it {Implications of the statement: no side effects were observed}},  {\em N Engl J Med} {\bf 292} (1975) 372--373.

\bibitem{woit1985topcharge}
P.~WOIT, {\em TOPOLOGICAL CHARGE IN LATTICE GAUGE THEORY}.
\newblock PhD thesis, 1985, 1985.
\newblock Copyright - Database copyright ProQuest LLC; ProQuest does not claim copyright in the individual underlying works; Last updated - 2023-02-19.

\bibitem{Alexandrou:2017hqw}
C.~Alexandrou, A.~Athenodorou, K.~Cichy, A.~Dromard, E.~Garcia-Ramos, K.~Jansen, U.~Wenger, and F.~Zimmermann, {\it {Comparison of topological charge definitions in Lattice QCD}},  {\em Eur. Phys. J. C} {\bf 80} (2020), no.~5 424, [\href{http://arxiv.org/abs/1708.00696}{{\tt arXiv:1708.00696}}].

\bibitem{Anber:2022qsz}
M.~M. Anber and E.~Poppitz, {\it {The gaugino condensate from asymmetric four-torus with twists}},  {\em JHEP} {\bf 01} (2023) 118, [\href{http://arxiv.org/abs/2210.13568}{{\tt arXiv:2210.13568}}].

\bibitem{GarciaPerez:2000aiw}
M.~Garcia~Perez, A.~Gonzalez-Arroyo, and C.~Pena, {\it {Perturbative construction of selfdual configurations on the torus}},  {\em JHEP} {\bf 09} (2000) 033, [\href{http://arxiv.org/abs/hep-th/0007113}{{\tt hep-th/0007113}}].

\bibitem{GARCIAPEREZ:1993}
M.~{García Pérez}, A.~González-Arroyo, P.~Martńez, L.~Fernández, A.~Sudupe, J.~Ruiz-Lorenzo, V.~Azcoiti, I.~Campos, J.~Ciria, A.~Cruz, D.~Iñiguez, F.~Lesmes, C.~Piedrafita, A.~Rivero, A.~Tarancón, P.~Téllez, D.~Badoni, and J.~Pech, {\it Instanton-like contributions to the dynamics of yang-mills fields on the twisted torus},  {\em Physics Letters B} {\bf 305} (1993), no.~4 366--374.

\end{thebibliography}\endgroup
 
  \bibliographystyle{JHEP}

\end{document}